\documentclass[epj]{svjour}

\usepackage{graphicx,amscd,amsmath,amssymb,verbatim}
\usepackage[dvips]{hyperref}
\usepackage[TS1,OT1,T1]{fontenc}

\begin{document}

\title{Poincar\'e and Heisenberg quantum dynamical symmetry: Casimir
invariant field equations of the quaplectic group }
\author{Stephen G. Low}
\institute{Stephen.Low@hp.com}
\date{Feb 5, 2005}
\abstract{
The unitary irreducible representations of a Lie group defines the
Hilbert space on which the representations act.\ \ If this Lie group
is a physical quantum dynamical symmetry group, this Hilbert space
is identified with the physical quantum state space. Hermitian representation
of the algebra are observables. The eigenvalue equations for the
representation of the set of Casimir invariant operators define
the field equations of the system.\ \ The Poincar\'e\ \ group is
the archetypical example with the unitary representations defining
the Hilbert space of relativistic particle states and the\ \ Klein-Gordon,
Dirac, Maxwell equations are obtained from the representations of
the Casimir invariant operators eigenvalue equations.\ \ The representation
of the Heisenberg group does not appear in this derivation. The
unitary representations of the Heisenberg group, however, play a
fundamental role in nonrelativistic quantum mechanics, defining
the Hilbert space and the basic momentum and position commutation
relations. Viewing the Heisenberg group as a generalized non-abelian
{\itshape translation} group, we look for a semidirect product group
with it as the normal subgroup that also contains the Poincar\'e\ \ group.
The quaplectic group, that is derived from a simple argument using
Born's orthogonal metric hypothesis, contains four Poincar\'e\ \ subgroups
as well as the normal Heisenberg subgroup. The general set of field
equations are derived using the Mackey representation theory for
general semidirect product groups. The simplest case of these field
equations is the relativistic-oscillator that plays a role in this
theory analogous to the Klein-Gordon equation in the Poincar\'e\ \ theory.
This theory requires a conjugate relativity principle that bounds
forces.\ \ Position-time space is no longer an invariant subspace
and the quaplectic transformations act on the full nonabelian time-position-momentum-energy
space with different observers measuring different position-time
subspaces.
}

\authorrunning{S. G. Low}
\titlerunning{Poincar\'e and Heisenberg quantum dynamical symmetry: The quaplectic group}
\maketitle
\section{Introduction}

The unitary irreducible representations of a Lie group are unitary
operators on a Hilbert space defining transitions between states
and the representation of the algebra are Hermitian operators that
are observables. The eigenvalue equations for the representations
of the set of Casimir invariant operators define the field equations
of the system.\ \ The Hilbert space is not given {\itshape a priori},
but rather is determined by the unitary representations of the group.
We say the Lie group is a dynamical symmetry group if this Hilbert
space may be identified with the physical quantum state space with
the field equations determining the physical single particle states.

 The Poincar\'e group of special relativity is the archetype dynamical
symmetry group. The Poincar\'e group acts naturally on four dimensional
position-time space. Time is not an invariant subspace under the
action of this group and is observer\ \ dependent. The Poincar\'e
group is a semidirect product of the cover of the Lorentz group
and the translation group and its unitary irreducible representations
may be obtained from the Mackey theory for semidirect product groups.\ \ The
field equations defined by the representations define the basic
single particle equations of physics; Klein-Gordon, Dirac, Maxwell
and so forth. The eigenvalues of the representations of the Casimir
operators are spin or helicity and mass.\ \ 

The Heisenberg group is the semidirect product of two translation
groups. It may be regarded as the generalization of a translation
group acting on a non-abelian phase space of position and momentum.
Again, its representations may be obtained from the Mackey representation
theory. The Hilbert space of basic non-relativistic quantum mechanics
is defined by the unitary representations of the Heisenberg group.
The Lie algebra are the Heisenberg commutators of position and momentum.
There is a single Casimir invariant that is the center of the algebra
and the field equations of this group are therefore trivial.

It is quite remarkable that the representations of the Poincar\'e
group gives rise to basic equations for single particle states without
any reference to the Heisenberg group that is perceived to be fundamental
to quantum mechanics. 

The question that this paper addresses is: What is the consequences
of defining a group that encompasses both the Poincar\'e and Heisenberg
groups.\ \ 

The group satisfying this property is a semi-direct product with
a generalized {\itshape translation} normal subgroup $\mathcal{N}$
that is the Heisenberg group $ ($which itself is the semidirect
product of translation groups$ )$.\ \ The relevant automorphisms
of the Heisenberg group are the symplectic group. This together
with the requirement for an orthogonal metric leads to the pseudo-unitary
group for the homogeneous group $\mathcal{K}$.\ \ This group acts
on a nonabelian position, time, energy, momentum phase space.\ \ It
contains four Poincar\'e subgroups, two associated with special
$ ($velocity$ )$ relativity and two defining a similar relativity
principle that generalizes the concept of force to hyperbolic rotations
on the momentum-time and energy-position subspaces.\ \ We call this
group the {\itshape quaplectic} group. It is a subgroup of the inhomogeneous
symplectic group, contains four $ ($a quad of$ )$ Poincar\'e groups
and has the quantum Heisenberg {\itshape translations }as a normal
subgroup.\ \ \ The theory embodies Born's reciprocity\ \ \cite{born1},
\cite{born2} and the new relativity principle mentioned.

The physical meaning and motivation of the quaplectic group is presented
and the unitary irreducible representation of the group and the
Hermitian representations of the quaplectic Lie algebra are determined
using Mackey's theory of representations of semidirect products.
This is the same Mackey theory that is used to determine the unitary
irreducible representations of the Poincar\'e group \cite{mackey},
\cite{dawber}, \cite{Low}. As with the Poincar\'e group, the choice
of the group defines the Hilbert space that is identified with quantum
particle states for the representations in question. The field equations
that are the eigenvalue equations of the representation of the Casimir
operators, are obtained and investigated. 
\section{Dynamical groups in quantum mechanics }\label{Section dynamcial
groups}

The unitary irreducible representations $\varrho $ of\ \ a Lie group
$\mathcal{G}$ act as unitary operators on a Hilbert space $\text{\boldmath
$\mathrm{H}$}^{\varrho }$. That is, if\ \ $g\in \mathcal{G}$
\begin{equation}
\varrho ( g) :\text{\boldmath $\mathrm{H}$}^{\varrho }\rightarrow
\text{\boldmath $\mathrm{H}$}^{\varrho }:\left| \psi \right\rangle
\mapsto \varrho ( g) \left| \psi \right\rangle  %
\label{unitary action on hilbert space}
\end{equation}

\noindent where $\varrho ( g) ^{\dagger }=\varrho ( g) ^{-1}$.\ \ The
Hilbert space $\text{\boldmath $\mathrm{H}$}^{\varrho }$ is determined
by the group and the unitary irreducible representation $\varrho
$ and so we\ \ label it with the representation with the group label
implicit.\ \ The Lie algebra of the group may be identified with
the tangent space of the group $\text{\boldmath $\mathrm{a}$}( \mathcal{G})
\simeq T_{e}\mathcal{G}$\ \ and the lift to the representation $\varrho
^{\prime }$ of elements of the algebra $X\in \text{\boldmath $\mathrm{a}$}(
\mathcal{G}) $ as Hermitian operators on $\text{\boldmath $\mathrm{H}$}^{\varrho
}$ 
\[
\varrho ^{\prime }( X) :\text{\boldmath $\mathrm{H}$}^{\varrho }\rightarrow
\text{\boldmath $\mathrm{H}$}^{\varrho }:\left| \psi \right\rangle
\mapsto \varrho ^{\prime }( X) \left| \psi \right\rangle  
\]

\noindent with $\varrho ^{\prime }( X) ^{\dagger }=\varrho ^{\prime
}( X) $.\ \ The choice of Hermitian representations for the algebra
requires a comment. The natural definition is in terms of anti-Hermitian
operators that follows from the unitary condition. That is, if $g=
e ^{X}$, then
\[
\varrho ^{\prime }( g) ^{\dagger } =\left(   e ^{\left( X\right)
\varrho ^{\prime }}\right) ^{\dagger }= e ^{\left( \varrho ^{\prime
}( X) ^{\dagger }\right) }=\varrho ( g) ^{-1}= e ^{-\varrho ^{\prime
}( X) } .
\]

\noindent and so $\varrho ^{\prime }( X) ^{\dagger }=-\varrho ^{\prime
}( X) $.\ \ Defining the Hermitian operator $\tilde{\varrho }^{\prime
}( X) =i \varrho ^{\prime }( X) $,\ \ it follows that $\varrho (
g)  = e ^{-i \tilde{\varrho }^{\prime }( X) }$. Hence forth, we
drop the tilde and always uses Hermitian representations of the
algebra. Note that an immediate consequence is the appearance of
an $i$ in the Lie algebra of the Hermitian representation of the\ \ elements
$X,Y,Z \in \text{\boldmath $\mathrm{a}$}( \mathcal{G}) $ \cite{Hall}.\ \ That
is, if\ \ \ $[X,Y]=Z$ then\ \ $[\varrho ^{\prime }( X) ,\varrho
^{\prime }( Y) ]=i \varrho ^{\prime }( Z) $.\ \ 

Quantum mechanics defines particles states $|\psi \rangle $ to be
elements of a Hilbert space $\text{\boldmath $\mathrm{H}$}$. Observables
are Hermitian operators $H=H^{\dagger }$, $H:\text{\boldmath $\mathrm{H}$}\rightarrow
\text{\boldmath $\mathrm{H}$}:|\psi \rangle \mapsto H|\psi \rangle
$ and unitary operators $U^{\dagger }=U^{-1}$ defines the evolution
of the states. A complete orthonormal basis $|\varphi _{x}\rangle
$ satisfies $\langle \varphi _{\tilde{x}}|\varphi _{x}\rangle =\delta
_{\tilde{x},x}$ and\ \ $|\varphi _{x}\rangle \langle \varphi _{\tilde{x}}|=I$
where $I$ is the identity on $\text{\boldmath $\mathrm{H}$}$.\ \ Observable
matrix elements are $h_{\tilde{x},x}=\langle \varphi _{\tilde{x}}|H|\varphi
_{x}\rangle $\ \ with $h_{\tilde{x},x}\in \mathbb{R}$.\ \ A unitary
operator $U$ transforms orthonormal bases into orthonormal bases
$|\tilde{\varphi }_{x}\rangle =U|\varphi _{x}\rangle $ and hence
preserves probabilities $\langle \tilde{\varphi }_{\tilde{x}}|\tilde{\varphi
}_{x}\rangle =\langle \tilde{\varphi }_{\tilde{x}}|U U^{\dagger
}|\varphi _{x}\rangle =\langle \tilde{\varphi }_{\tilde{x}}|\varphi
_{x}\rangle $.\ \ The matrix elements of an observable $H$ are $\langle
\tilde{\varphi }_{\tilde{x}}|H\mathsf{|}\tilde{\varphi }_{x}\rangle
=\langle \tilde{\varphi }_{\tilde{x}}| U^{\dagger }U H U^{\dagger
}U|\varphi _{x}\rangle =\langle \tilde{\varphi }_{\tilde{x}}|\tilde{H}\mathsf{|}\tilde{\varphi
}_{x}\rangle $ with $\tilde{H}=U H U^{\dagger }$.

Casimir invariant operators $C_{\alpha }$\ \ with $\alpha =1,2,...N_{c}$
are elements of the enveloping algebra of the Lie algebra of the
group, $C_{\alpha }\in e( \mathcal{G}) $, that commute with all
elements of the algebra; $[C_{\alpha },X]=0$ for all $X\in \text{\boldmath
$\mathrm{a}$}( \mathcal{G}) $. The number of independent Casimir
invariant operators is $N_{c}=N_{g}-N_{r}$. $N_{g}$ is the dimension
of the Lie algebra and $N_{r}$ is its rank.\ \ The Hermitian irreducible
representation $\varrho ^{\prime }( C_{\alpha }) $ have eigenvalues
$c_{\alpha }$ that are real constants for a given unitary irreducible
representation 
\begin{equation}
\varrho ^{\prime }( C_{\alpha }) \left| \psi \right\rangle  =c_{\alpha
}\left| \psi \right\rangle   \mathrm{with} \left| \psi \right\rangle
\in \text{\boldmath $\mathrm{H}$}^{\varrho },\ \ \ \alpha =1,2...N_{c}.%
\label{field equation definition}
\end{equation}

These equations in the physical theory are the {\itshape field equations}
for the dynamical group. The simultaneous solution of these eigenvalue
equations define the observable particle states of the theory and
the eigenvalues define physically observable, constant properties
that are attributed to these particle states. 

We will say that $\mathcal{G}$ is a dynamical symmetry of a quantum
system if the following conditions are met.\ \ First, the Hilbert
space $\text{\boldmath $\mathrm{H}$}^{\varrho }$ determined by the
unitary irreducible representations $\varrho $ of a group $\mathcal{G}$
is identical to the Hilbert space $\text{\boldmath $\mathrm{H}$}$
of the quantum system in question, $\text{\boldmath $\mathrm{H}$}\simeq
\text{\boldmath $\mathrm{H}$}^{\varrho }$. This means that the $
($particle$ )$ states of the quantum system are states in the unitary
irreducible representations of the dynamical group.\ \ Unitary operators
$U$ defining unitary evolution of the system are given by $U=\varrho
( g) $ with corresponding observables $H=\varrho ^{\prime }( X)
$.\ \ The field equations defining the observable particle state
are defined by the eigenvalue equation for the representations of
the Casimir invariant operators. It is because these dynamical aspects
of the physics arise from the group that we call the group a dynamical
symmetry, rather than just a symmetry. This definition of a dynamical
group is motivated by the {\itshape spectrum generating} or {\itshape
dynamical} groups as defined by Bohm \cite{bohm} and reference there-in.

The dynamical groups of interest have the form of a semidirect product
Lie group $\mathcal{G}=\mathcal{K}\otimes _{s}\mathcal{A}$ with
$\mathcal{A}$ is the closed normal subgroup and $\mathcal{K}$ the
homogeneous group. Furthermore, the Lie groups of interest are real
matrix groups that are algebraic. That is, the groups are closed
subgroups of $\mathcal{G}\mathcal{L}( n,\mathbb{R}) $, that are
defined by polynomial, or algebraic, constraints.\ \ Examples of
groups of these type include the special orthogonal, symplectic,
unitary, Euclidean, Poincar\'e, Heisenberg groups as well as, will
be shown,\ \ the quaplectic group.\ \ 

The translation group $\mathcal{T}( n+1) $ acts on an abelian position-time
manifold $\mathbb{M}\simeq \mathbb{R}^{n+1}$ that may be identified
with the physical concept of a flat $n+1$ dimensional position-time
space or {\itshape space-time}. The usual physical case is $n=3$.

As noted, the Poincar\'e group $\mathcal{P}$ is the archetypical
dynamical group. It is the cover of the group $\mathcal{E}( 1,n)
=\mathcal{S}\mathcal{O}( 1,n) \otimes _{s}\mathcal{T}( n+1) $ acting
on a $n$+1 dimensional position-time space. Then,\ \ for the usual
physical case $n=3$, $\mathcal{P}=\overline{\mathcal{E}}( 1,3) $.
The quantum state space is the Hilbert space of the unitary irreducible
representations of the Poincar\'e group. The Mackey representation
theory shows that these Hilbert spaces are \label{RefMark:113145932}
\[
\text{\boldmath $H$}^{\varrho }= \text{\boldmath $H$}^{\sigma }\otimes
\text{\boldmath $L$}^{2}( \mathbb{A},\mathbb{C}) \ \ 
\]

\noindent where $\mathbb{A}\simeq \overline{\mathcal{S}\mathcal{O}}(
1,n) /\mathcal{K}\mbox{}^{\circ}$ and $\mathcal{K}\mbox{}^{\circ}\simeq
\overline{\mathcal{S}\mathcal{O}}( n) $,\ \ $\overline{\mathcal{E}}(
n-1) $, or\ \ $\overline{\mathcal{S}\mathcal{O}}( 1,n-1) $ depending
on whether the representations are timelike, null or spacelike.
The Hilbert space\ \ $\text{\boldmath $H$}^{\sigma }$ is the corresponding
Hilbert space of the unitary irreducible representation of the little
group $\mathcal{K}\mbox{}^{\circ}$. For the timelike case, these
are finite dimensional and for the physical case $n=3$, it is just
the $2j+1$ dimensional representation spaces of $\mathcal{S}\mathcal{U}(
2) $ with $j$ half integral.\ \ Note that the symmetric spaces $\mathbb{A}$
are the timelike, null and spacelike hyperboloids \cite{Low}

There are two Casimir invariant operators in this case with eigenvalue
equations for the representations of the Casimir operators acting
on the quantum state space of the form $ ($2$ )$. The two eigenvalues
may be associated with the physical concepts of mass and spin $
($or helicity$ )$ that are constants for each of the irreducible
representations. Solution of these eigenvalue equations for the
various irreducible representations defines the Klein-Gordon, Dirac,
Maxwell and so forth, field equations \cite{dawber}. 
\section{ The quaplectic dynamical group }\label{c:20}\label{Groups
in QM}

A dynamical group may also act on the a nonabelian phase space.
The Heisenberg group $\mathcal{H}( n) =\mathcal{T}( n) \otimes _{s}\mathcal{T}(
n+1) $ acts as a nonabelian ``translation'' group on a $2n$ dimensional
position-momentum phase space.\ \ More generally, $\mathcal{H}(
n+1) $ acts on the nonabelian $2n+2$ dimensional position-time-momentum-energy
phase space. 

In this section we determine the simplest semidirect product group
that contains the Poincar\'e group as a subgroup and also has the
Heisenberg group as a normal subgroup. We call the group obtained
the {\itshape quaplectic} group.\ \ We then\ \ examine the consequences
of it acting as a dynamical symmetry group by determining its unitary
irreducible representations and the associated Casimir field equations.\ \ 

The question investigated in this paper is the group $\mathcal{G}=\mathcal{K}\otimes
_{s}\mathcal{H}(n+1)$ that acts as a dynamical group on the nonabelian
phase space analogous to the action of the group $\mathcal{E}( 1,n)
=\mathcal{S}\mathcal{O}( 1,n) \otimes _{s}\mathcal{T}( n+1) $ on
the abelian position-time space. 
\subsection{Group properties}\label{c:21}

Consider a semidirect product group $\mathcal{G}=\mathcal{K}\otimes
_{s}\mathcal{H}(n+1)$. The first question is to determine the conditions
on a subgroup $\mathcal{K}$ that are required in order that this
semidirect product can be constructed. As $\mathcal{H}( n+1) $ is
a normal subgroup of $\mathcal{G}$, $\mathcal{K}$ must be a subgroup
of the automorphisms of\ \ $\mathcal{H}( n+1) $. The group product
for the elements of the Heisenberg group $h( w,\iota ) \in \mathcal{H}(
n+1) $ may be written
\begin{equation}
\begin{array}{l}
 h( \tilde{w,}\tilde{\iota })  \cdot h( w,\iota ) = h( w+\tilde{w},
\iota +\tilde{\iota }+{}\tilde{w}\cdot \zeta \cdot w)  \\
 h^{-1}( w,\iota ) =g( -w,-\iota ) 
\end{array}%
\label{XRef-Equation-113145932}
\end{equation}

\noindent where $g( 0,0) $ is the identity element and $w\in \mathbb{R}^{2(n+1)}$,
$\iota \in \mathbb{R}$ and ${}\tilde{w}\cdot \zeta \cdot w={}\tilde{w}^{a}\zeta
_{a,b}w^{b}$ with the indices $a,b,..=0,1,...n$ and $i,j,..=1,...n$.
This index convention is always assumed to be the case unless explicitly
noted otherwise.\ \ \ The components\ \ of the\ \ symplectic metric
$\zeta $ in these canonical coordinates is the\ \ $2(n+1)\times
2(n+1)$ matrix
\[
\zeta =\left( \begin{array}{ll}
 0 & I \\
 -I & 0
\end{array}\right) ,
\]

\noindent where $I$ is $(n+1)\times (n+1)$ identity matrix. Therefore,
while the Heisenberg group has no implicit concept of an orthogonal
metric,\ \ it does have implicit in its definition a symplectic
structure. The Heisenberg group is a semidirect product $\mathcal{H}(
n+1) \simeq \mathcal{T}( n+1) \otimes _{s}\mathcal{T}( n+2) $ and\ \ it
is a matrix group that is algebraic 
\[
h( w,\iota ) \simeq \left( \begin{array}{lll}
 I & 0 & w \\
 {}w\cdot \zeta  & 1 & \iota  \\
 0 & 0 & 1
\end{array}\right) .
\]

The Lie algebra is spanned by the basis $\{W_{\mu }\}$ with $\mu
,\nu =0,...n, \tilde{0},...\tilde{n}$ that satisfies the Lie algebra
relations 
\[
\left[ W_{\mu },W_{\nu }\right] =\zeta _{\mu ,\nu }I.
\]

\noindent If we identify $\{W_{\mu }\}=\{T,P_{i},-E,Q_{i}\}$ then\ \ $[P_{i},Q_{j}]
= \delta _{i,j}I $ and $[T,E]=-I$.\ \ $ ($We are using natural units
with $\hbar =1$.\ \ Units are discussed further shortly.$ )$

The action $ \varsigma _{a}h\doteq a\cdot h\cdot a^{-1}$ of the
Heisenberg group on itself are the automorphisms 
\begin{equation}
\varsigma _{h( \tilde{w},\tilde{\iota }) }h( w,\iota ) = h( w, \iota
+2{}w\cdot \zeta \cdot \tilde{w}) ,%
\label{group automorphism for automorphism group}
\end{equation}

\noindent Elements of the complete group of linear automorphisms
\cite{folland} of $\mathcal{H}( n+1) $ have the action
\begin{equation}
\varsigma _{a_{\pm }( \tilde{\varepsilon },\tilde{A},\tilde{w},\tilde{\iota
}) }h( w,\iota ) = h( \tilde{\varepsilon } \tilde{A}w, \pm \varepsilon
^{2}( \iota +2{}w\cdot \zeta \cdot \tilde{w}) ) ,%
\label{group automorphism for automorphism group}
\end{equation}

\noindent where $a( 1,A,0,0) \in \mathcal{S}p( 2n+2) $, $a( \epsilon
,I,0,0) \in \mathcal{A}b( 1) $ is the one parameter, $\epsilon 
$, abelian group and the discrete symmetry is $a_{\pm }\in \mathcal{D}_{2}$.
Thus, the group of\ \ linear automorphisms \cite{folland} $\mathcal{A}ut(
n+1) $ of $\mathcal{H}( n+1) $has the form
\[
\begin{array}{rl}
 \mathcal{A}ut( n+1)  & =\mathcal{D}_{2}\otimes \left( \mathcal{A}b(
1) \otimes \mathcal{S}p( 2n+2) \right) \otimes _{s}\mathcal{H}(
n+1)  \\
  & =\mathcal{D}_{2}\otimes \mathcal{A}b( 1) \otimes _{s}\mathcal{H}\mathcal{S}p(
n+1) 
\end{array}
\]

\noindent where $\mathcal{H}\mathcal{S}p( n) \doteq \mathcal{S}p(
2n) \otimes _{s}\mathcal{H}( n) $. A general element of the continuous
automorphisms $\mathcal{A}ut( n+1) $ may be written
\[
a( \varepsilon ,A,w,\iota ) =a( \varepsilon ,I,0,0) \cdot a( 1,A,0,0)
\cdot a( 1,I,w,\iota ) 
\]

\noindent and a general element $a( \varepsilon ,A,w,\iota ) $ is
represented by the matrix group 
\[
a( \varepsilon ,A,w,\iota ) \simeq \left( \begin{array}{lll}
 A & 0 & A\cdot w \\
 \varepsilon  w\cdot \zeta  & \varepsilon  & \varepsilon  \iota
\\
 0 & 0 & \varepsilon ^{-1}
\end{array}\right) ,
\]

\noindent where\ \ $\varepsilon \in \mathbb{R}\backslash \{0\}$
and\ \ $A\in \mathcal{S}p( 2n+2) \text{}$defined by $A_{1}..A_{4}$
real $(n+1)\times (n+1)$ matrices satisfying
\[
A=\left( \begin{array}{ll}
 {}A_{1} & {}A_{3} \\
 {}A_{4} & {}A_{2}
\end{array}\right) ,\ \ A^{-1}=\left( \begin{array}{ll}
 {}^{t}A_{2} & -{}^{t}A_{3} \\
 -{}^{t}A_{4} & {}^{t}A_{1}
\end{array}\right) .
\]

The group composition law and inverse may be computed directly from
this faithful matrix representation, or abstractly by using the
property Heisenberg group is a normal subgroup and therefore\ \ 
\begin{equation}
\begin{array}{l}
 a( \tilde{\varepsilon },\tilde{A},\tilde{w},\tilde{\iota }) \cdot
a( \varepsilon ,A,w,\iota )  \\
 =a( \tilde{\varepsilon },\tilde{A}) \cdot a( \varepsilon ,A) \cdot
a( \varepsilon ,A) ^{-1}\cdot h( \tilde{w},\tilde{\iota }) \cdot
a( \varepsilon ,A) \cdot h( w,\iota )  \\
 =a( \tilde{\varepsilon }\varepsilon , \tilde{A}\cdot A,w+\varepsilon
^{-1}A^{-1}\cdot \tilde{w}, \iota +\varepsilon ^{-2}\tilde{\iota
}+\varepsilon ^{-1}{}\tilde{w}\cdot \zeta \cdot A\cdot w) 
\end{array}%
\label{group mulitplication for automorphism group}
\end{equation}

The semidirect product group $\mathcal{G}$ with $\mathcal{H}( n+1)
$ as a normal subgroup must, in general, be a subgroup of $\mathcal{A}ut(
n+1) $. In this initial investigation, we do not consider the effects
of $\mathcal{A}b( 1) $ nor the discrete automorphisms further and
so consider $\mathcal{G}$ to be a subgroup of $\mathcal{H}\mathcal{S}p(
n+1) $.\ \ Elements $g\text{}$ of\ \ the\ \ $\mathcal{H}\mathcal{S}p(
n+1) $ subgroup are $g( A,w,\iota ) \simeq a( 1,A,w,\iota ) $.\ \ \ 

Consider the particular automorphism $a( 1,\Upsilon \mbox{}^{\circ},0,0)
\in \mathcal{A}ut( n+1) $, and therefore $\Upsilon \mbox{}^{\circ}\in
\mathcal{S}p( 2n+2)  $ that is the particular element that has the
property $\hat{w}=\Upsilon \mbox{}^{\circ}\cdot  w$ with\ \ $\hat{w}=(-w^{\tilde{0}},w^{1},..w^{n},w^{0},w^{\tilde{1}}...w^{\tilde{n}})$.
Applying the automorphism to the full $\mathcal{H}\mathcal{S}p(
n+1) $ group puts the components of the symplectic metric into the
following form with the associated symplectic inverse condition
\begin{equation}
\hat{\zeta }=\left( \begin{array}{ll}
 0 & \eta  \\
 -\eta  & 0
\end{array}\right) ,\ \ \hat{A}^{-1}=\left( \begin{array}{ll}
 \eta \cdot {}^{t}\hat{A}_{2}\cdot \eta  & -\eta \cdot {}^{t}\hat{A}_{3}\cdot
\eta  \\
 -\eta \cdot {}^{t}\hat{A}_{4}\cdot \eta  & \eta \cdot {}^{t}\hat{A}_{1}\cdot
\eta 
\end{array}\right)  .%
\label{symplect inverse with eta}
\end{equation}

\noindent where $\eta $ is $(n+1)\times (n+1)$ diagonal matrix with
diagonal $(-1,1,..1)$ and $\eta ^{-1}=\eta $. Note that the $\hat{A}_{\alpha
}$ are an intertwined combination of the components of the $A_{\alpha
}$.\ \ 

The Lie algebra is
\[
\left[ \hat{W}_{\mu },\hat{W}_{\nu }\right] =\hat{\zeta }_{\mu ,\nu
}I,\ \ \mu ,\nu =0,...n, \tilde{0},...\tilde{n}
\]

\noindent where now we identify $\{\hat{W}_{\mu }\}=\{E,P_{i},T,Q_{i}\}$.
Then, again,\ \ $[P_{i},Q_{j}] = \delta _{i,j}I $ and $[T,E]=-I$.\ \ That
the algebra is in, fact identical to that of the $\{W_{\mu }\}$
, emphasizes again that this group is isomorphic to the original
group as required under the action of the automorphism $a( 1,\Upsilon
\mbox{}^{\circ},0,0) $.\ \ Again, it is emphasized that $\mathcal{H}\mathcal{S}p(
n+1) $\ \ has no concept of an orthogonal metric and that "$\text{}\mathcal{H}\mathcal{S}p(
1,n) \simeq \mathcal{H}\mathcal{S}p( n+1) $". We use this form of
the $\mathcal{H}\mathcal{S}p( n+1) $ group and drop the tilde in
what follows. 

While $\mathcal{H}\mathcal{S}p( n+1) $ has no concept of an orthogonal
metric, the Poincar\'e group acting on the position-time and the
Poincar\'e group acting on energy-momentum spaces\ \ have\ \ Casimir
invariant operators that define pseudo-orthogonal metric structures
\[
 -\mathit{T}^{2}+\frac{1}{c^{2}}\mathit{Q}^{2} ,\ \ \ \ \ \ \ \ \ \ \ \ \ \ \ \mathit{P}^{2}-\frac{1}{c^{2}}\mathit{E}^{2}.
\]

Born \cite{born1},\cite{born2}\ \ based on a reciprocity principle,
argued that these metrics combine into a single metric on $2(n+1)$
dimensional phase space 
\begin{equation}
 -\mathit{T}^{2}+\frac{1}{c^{2}}\mathit{Q}^{2}+\frac{1}{b^{2}}\left(
\mathit{P}^{2}-\frac{1}{c^{2}}\mathit{E}^{2}\right) .%
\label{b definition}
\end{equation}

\noindent $b$ is a new universal dimensional physical constant with
units of force that we shall discuss shortly.\ \ The group with
both an orthogonal and symplectic structure is the unitary group
$\mathcal{U}( 1,n) \simeq \mathcal{O}( 2,2n) \cap \mathcal{S}p(
2n+2) $.\ \ This group leaves invariant the symplectic structure
as well as the orthogonal metric. 

As $\mathcal{U}( 1,n) $ is a subgroup of\ \ $\mathcal{S}p( 2n+2)
$ we may define a semidirect product\ \ group $\mathcal{Q}( 1,n)
$ that\ \ has a dimension\ \ $N=(n+2)^{2}$, that we name the quaplectic
group, by\ \ 
\begin{equation}
\begin{array}{rl}
 \mathcal{Q}( 1,n)  & =\mathcal{U}( 1,n) \otimes _{s}\mathcal{H}(
n+1)  \\
  & =\mathcal{S}\mathcal{U}( 1,n) \otimes _{s}\mathcal{O}s( n+1)
\end{array}
\end{equation}

\noindent where $\mathcal{O}s( n) =\mathcal{U}( 1) \otimes _{s}\mathcal{H}(
n) $. The orthogonal condition of $\mathcal{O}( 2,2n) $ implies
that, in addition to the symplectic inverse condition, $ ($3.7$
)$, the $A$ must also satisfy the\ \ condition 
\[
A^{-1}=\left( \begin{array}{ll}
 \eta \cdot {}^{t}A_{1}\cdot \eta  & \eta \cdot {}^{t}A_{4}\cdot
\eta  \\
 \eta \cdot {}^{t}A_{3}\cdot \eta  & \eta \cdot {}^{t}A_{2}\cdot
\eta 
\end{array}\right) 
\]

\noindent from which it follows that $A_{1}=A_{2}=\Lambda $ and
$A_{3 }=-A_{4}=M$ and imposes the conditions that\ \ $\Lambda ^{-1}=\eta
\cdot {}^{t}\Lambda \cdot \eta $ and\ \ ${}M^{-1}=-\eta \cdot {}^{t}M\cdot
\eta  $ and so the element of a faithful matrix representation of
$\mathcal{H}( n+1) $\ \ has the form
\[
g( \Lambda ,M,x,y,\iota ) \simeq \left( \begin{array}{llll}
 \Lambda  & M & 0 & \Lambda \cdot x+M\cdot y \\
 -M & \Lambda  & 0 & -M\cdot x+\Lambda \cdot y \\
 -y & x & 1 & \iota  \\
 0 & 0 & 0 & 1
\end{array}\right) 
\]

\noindent where $w=(x,y)$ and $x,y\in \mathbb{R}^{n+1}$.\ \ \ The
group multiplication law for $\mathcal{Q}(1,n$$ )$ may be written
in a complex notation by defining\ \ $z=x+i y $, $z\in \mathbb{C}^{n+1}$
and $\Upsilon =\Lambda +i M$.\ \ Then 
\begin{equation}
\begin{array}{l}
 g( \tilde{\Upsilon },\tilde{z},\tilde{\iota })  \cdot g( \Upsilon
,z,\iota )  \\
 = g( \tilde{\Upsilon }\cdot \Upsilon ,\tilde{z}+\Upsilon \cdot
z, \iota +\tilde{\iota }+\frac{i}{2}\left( \overline{z}\cdot \eta
\cdot \tilde{z}-\overline{\tilde{z}}\cdot \eta \cdot z\right) )
, \\
 g^{-1}( \Upsilon ,z,\iota ) =g( \Upsilon ^{-1},-\Upsilon ^{-1}\cdot
z,-\iota ) .
\end{array}
\end{equation}

\noindent An element $g\in \mathcal{Q}( 1,n) $ may be realized by
the $(n+1)\times (n+1)$ complex\ \ matrices 
\[
g( \Upsilon ,z,\iota ) \simeq \left( \begin{array}{lll}
 \Upsilon  & 0 & \Upsilon \cdot z \\
 \overline{z} & 1 & \iota  \\
 0 & 0 & 1
\end{array}\right) .
\]

\noindent The\ \ inner automorphisms of the group are 
\[
\varsigma _{g( \tilde{\Upsilon },\tilde{z},\tilde{\iota }) }g( \Upsilon
,z,\iota ) \simeq g( \tilde{\Upsilon }\cdot \Upsilon \cdot \tilde{\Upsilon
}^{-1},\tilde{z}+\tilde{\Upsilon }\cdot z, \iota +\iota +i( \overline{\tilde{z}}\cdot
\eta \cdot \tilde{z}-\overline{z}\cdot \eta \cdot \tilde{z}) ) .
\]

\noindent If $M=0$, we obtain the $\mathcal{S}\mathcal{O}( 1,n)
\otimes _{s}\mathcal{H}( n+1) $ subgroup with elements of the form
$g( \Lambda ,z,\iota ) $.
\subsection{Lie algebra and Casimir invariant operators}\label{c:23}

A general element of the Canonical algebra is $Z+A$ with $Z$ and
element of\ \ the algebra of $\mathcal{U}( 1,n) $ and $A$ an element
of the Heisenberg algebra of $\mathcal{H}( n+1) $,
\[
\begin{array}{l}
 Z=\phi ^{a,b}M_{a,b}+\varphi ^{a,b}L_{a,b}, \\
 A= \iota  I + x^{a}X_{a}+ y^{a}Y_{a}.
\end{array}
\]

\noindent with the parameters all real. These generators satisfy\ \ 
\begin{equation}
\begin{array}{l}
 \left[ L_{a,b},L_{c,d}\right] =-L_{b,d} \eta _{a,c}+L_{b,c} \eta
_{a,d}+L_{a,d} \eta _{b,c}-L_{a,c} \eta _{b,d} \\
 \left[ L_{a,b},M_{c,d}\right] =-M_{b,d} \eta _{a,c}-M_{b,c} \eta
_{a,d}+M_{a,d} \eta _{b,c}+M_{a,c} \eta _{b,d} \\
 \left[ M_{a,b},M_{c,d}\right] =-L_{b,d} \eta _{a,c}-L_{b,c} \eta
_{a,d}-L_{a,d} \eta _{b,c}-L_{a,c} \eta _{b,d} \\
 \left[ L_{a,b},X_{c}\right]  = -X_{b} \eta _{a,c}+X_{a} \eta _{b,c}
\\
 \left[ L_{a,b},Y_{c}\right]  = -Y_{b} \eta _{a,c}+Y_{a} \eta _{b,c}
\\
 \left[ M_{a,b},X_{c}\right]  = -Y_{b} \eta _{a,c}-Y_{a} \eta _{b,c}
\\
 \left[ M_{a,b},Y_{c}\right]  = X_{b} \eta _{a,c}+X_{a} \eta _{b,c}
\\
 \left[ X_{a},Y_{b}\right]  =I \eta _{a,b}
\end{array}%
\label{canonical real algebra}
\end{equation}

\noindent Clearly both the generators $\{L_{a,b},X_{c }\}$ and $\{L_{a,b},Y_{c
}\}$ define the algebras of Poincar\'e subgroups. The second order
Casimir invariant operator is
\[
C_{2}=\frac{1}{2}\eta ^{a,b}( X_{a}X_{b}+Y_{a}Y_{b}) +\frac{1}{2}\left(
n+1\right)  I - I U.
\]

As $I=C_{1}$ is a Casimir operator, and any linear combinations
of Casimir operators is\ \ also one, this term may be dropped.\ \ The
algebra of $\mathcal{Q}( 1,n) =\mathcal{U}( 1,n) \otimes \mathcal{H}(
n+1) $ may also be written in a complex form by defining
\[
A_{a}^{\pm }=\frac{1}{\sqrt{2}} \left( X_{a}\mp  i Y_{a}\right)
,\ \ \ Z_{a,b}= \frac{1}{2}\left( M_{a,b}- i L_{a,b}\right) .
\]

\noindent It follows directly that the Lie algebra relations take
the more condensed form
\begin{equation}
\begin{array}{l}
 \left[ Z_{a,b},Z_{c,d}\right]  =i(  Z_{c,b} \eta _{a,d}-Z_{a,d}
\eta _{b,c}) , \\
 \begin{array}{ll}
 \left[ A_{a}^{+},A_{b}^{-}\right] =i \eta _{a,b}\mathit{I}\mathit{,}
& \left[ Z_{a,b},A_{c}^{\pm }\right] =\mp i \eta _{a,c}A_{b}^{\pm
}
\end{array}
\end{array}%
\label{Canonical Algebra}
\end{equation}

\noindent where as usual, the indices $a,b=0,1,...n$.\ \ $\{Z_{a,b}\}$
span the algebra of $\mathcal{U}( 1,n) $ and $\{A_{c}^{\pm },I\}$
span the algebra of $\mathcal{H}( 1,n) $.

The compact case of $\mathcal{Q}( n) =\mathcal{U}( n) \otimes \mathcal{H}(
n) $ is immediately obtained by restricting the\ \ indices $a,b$
to $i,j=1,..n\text{}$. The\ \ \ generators $\{Z_{i,j},A_{k}^{\pm
},I\}$\ \ \ span the\ \ algebra of the subgroup\ \ $\mathcal{Q}(
n) $ of $\mathcal{Q}( 1,n) $, where we note that\ \ \ $\eta _{i,j}=\delta
_{i,j}$. 

Define $U\doteq  \eta ^{a,b}Z_{a,b}$. This is the $\mathcal{U}(
1)  $ generator in the decomposition $\mathcal{Q}( 1,n) =\mathcal{S}\mathcal{U}(
1,n) \otimes _{s}\mathcal{O}s( 1,n) $ with $\mathcal{O}s( 1,n) =\mathcal{U}(
1) \otimes _{s}\mathcal{H}( 1,n) $.\ \ The corresponding generators
of the algebra of $\mathcal{S}\mathcal{U}( 1,n) $ are\ \ $\{\hat{Z}_{a,b}\}$
\begin{equation}
Z_{a,b}= \hat{Z}_{a,b}+\frac{1}{n+1}U \eta _{a,b }%
\label{canonical algebra su}
\end{equation}

\noindent Note that $\eta ^{a,b}\hat{Z}_{a,b}=0$\ \ and that the
$Z_{a,b}$ and $\hat{Z}_{a,b}$ commute with $U$ and the $\hat{Z}_{a,b}$
satisfy the same commutation relations as defined above in $ ($12$
)$.

For $\mathcal{Q}( 1,n) $, $N_{c}=n+2$ and the $n+2$ Casimir invariant
operators are \cite{quesne}
\begin{equation}
\begin{array}{l}
 C_{1}= I,\ \ \ \ \ \ C_{2}=\eta ^{a_{1},a_{2}}W_{a_{1},a_{2}} ,\ \ \ \ \ ...
\\
 C_{2\beta }=\eta ^{a_{1},a_{2\beta }}...\eta ^{a_{2\beta -2},a_{2\beta
-1}}W_{a_{1},a_{2}}...W_{a_{2\beta -1},a_{2\beta }},
\end{array}%
\label{Casimir invariants of cannonical group in complex basis}
\end{equation}

\noindent with $\beta =1,.n+1$ and\ \ $W_{a,b}\doteq  A^{+}_{a}A^{-}_{b}-I
Z_{a,b}$.\ \ Note that the second order invariant is of the form
\begin{equation}
C_{2}= \eta ^{a,b}A^{+}_{a}A^{-}_{b}-I U=A^{2}-I U,%
\label{canonical algebra casimir second order simple form}
\end{equation}

\noindent where $U$ is the generator of the algebra of $ \mathcal{U}$$
($1$ )$ defined above. The commutation relations for the $W_{a,b}$
are 
\begin{equation}
\begin{array}{l}
 \left[ Z_{a,b},W_{c,d}\right]  =i( \eta _{a,d}W_{b,c}-\eta _{b,c}W_{d,a})
\\
  \left[ A_{c}^{\pm },W_{a,b}\right] =0 
\end{array}
\end{equation}

\noindent and therefore $W_{c,d}$ are Heisenberg {\itshape translation}
invariant. It is important to note that both of the terms in $W_{a,b}$
are required in order for the commutator to vanish with $A_{c}^{\pm
}$.\ \ The $W_{c,d}$ commutators with $Z_{a,b}$ are the same as
$Z_{c,d}$.\ \ The Casimir invariants of $\mathcal{U}( 1,n) $ are
\cite{popov}
\begin{equation}
D_{\beta }=\eta ^{a_{1},a_{2\beta }}...\eta ^{a_{2\beta -2},a_{2\beta
-1}}Z_{a_{1},a_{2}}...Z_{a_{2\beta -1},a_{2\beta }}, %
\label{Unitary Casimir Invariants}
\end{equation}

\noindent where $\beta =1,...,n+1.$ Therefore, $ ($3.14$ )$ are
invariant unitary $\mathcal{U}( 1,n) $ rotations and, as the $W_{a,b}$
have already been established to be Heisenberg translational invariant,
it follows that they are\ \ Casimir invariants of $\mathcal{Q}(
1,n) $.\ \ Note also that it follows immediately that\ \ 
\begin{equation}
\left[ D_{\beta },D_{\alpha }\right] =0, \left. 
\text{$[D_{\beta },I]=0$,} [D_{\beta },C_{2\alpha }\right] =0 ,%
\label{Quaplectic commute}
\end{equation}

\noindent with $ \alpha ,\beta =1,...n+1$.\ \ As an aside that will
be useful, we note also that $\mathcal{O}s( n) =\mathcal{U}( 1)
\otimes _{s}\mathcal{H}( n) $ is\ \ rank 2 for all $n$.\ \ The\ \ two
Casimir invariants $\tilde{\mathit{C}}_{1}$\ \ and $\tilde{\mathit{C}}_{2}$\ \ are\ \ 
\begin{equation}
\tilde{\mathit{C}}_{1}=C_{1}=I,\ \ \ \ \ \ \tilde{\mathit{C}}_{2}=C_{2}=A^{2}-
I U.%
\label{Casimir invariants oscillator group}
\end{equation}
\subsection{Physical interpretation of the quaplectic group}

The Poincar\'e group is a dynamical symmetry of Minkowski space
$\mathbb{M}^{1,n}=\mathcal{E}( 1,n) /\mathcal{S}\mathcal{O}( 1,n)
$.\ \ The generators $\{Y_{a}\}=\{E,P_{i}\}$ of the translation
group on position-time space are identified with energy and momentum.\ \ In
the quantum theory, the unitary representations of the Poincar\'e
group are required and the corresponding Hermitian representation
of these generators define the observable energy and momentum degrees
of freedom.

The quaplectic group acts on the nonabelian phase space $\mathbb{Q}^{1,n}=\mathcal{Q}(
1,n) /\mathcal{S}\mathcal{U}( 1,n) $.\ \ Again the generators $\{Y_{a}\}$\ \ are
the energy and momentum but the Heisenberg algebra also includes
the\ \ $\{X_{a}\}=\{T,Q_{i}\}$ that are the time and position degrees
of freedom. Again, in the quantum theory, the Hermitian representations
of these generators defines these observable degrees of freedom.

Macroscopic physics assigns different dimensions to the time, energy,
position and momentum degrees of freedom that are usually defined
in terms of the constants $c,\hbar $ and $G$.\ \ Instead of $G$,
we choose to use the constant $b$ introduced in Section $ ($8$ )$
that has units of force and $G$ may be defined in terms of it as
$G=\alpha _{G}c^{4}/b$ where $\alpha _{G}$ is just the dimensionless
gravitational coupling constant to be determined by theory or experiment.\ \ Planck
scales $\lambda _{\alpha }$ of time, position, momentum and energy
may be defined in terms of these three constants as
\[
\lambda _{t}=\sqrt{\hbar /b c}, \lambda _{q}=\sqrt{\hbar  c/b},
\lambda _{p}=\sqrt{\hbar  b/c},\ \ \lambda _{e}=\sqrt{\hbar  b c}
\]

If $\alpha _{G}=1$, these are just the usual Planck scales. The
Lorentz subgroup of the Poincar\'e group transforms position and
time degrees of freedom into each other. This mixing is {\itshape
real} in the sense that observers in a frame measure {\itshape position}
and measure {\itshape time} of the position-time space differently
with the usual time dilation and length contraction effects when
measurements are compared between frames. The constant $c$ is required
to convert units of time into position and vice versa. This is no
different that if we measured the $x$ direction in meters and the
$y$ direction in feet so that every time a rotation was applied,
a constant converting meters into feet and vice versa is required.
Simply by choosing the right units, this constant is eliminated
and so it is also with $c$, in {\itshape natural} units $c=1$. 

The quaplectic group transforms all of the degrees of freedom, position,
time, energy and momentum into each other. A special case of this
is the usual Lorentz transformations of special $ ($velocity$ )$
relativity acting on the momentum-energy and position-time subspaces.
The full set of transformations cause a mixing of all the degrees
of freedom that are as {\itshape real} as the Lorentz velocity subset.
This mixing is {\itshape real} in the sense that observers in a
frame measure {\itshape position},\ \ {\itshape time, energy }and{\itshape
momentum} of the\ \ nonabelian\ \ position-time-energy-momentum
space differently with generalizations of the usual time dilation
and length contraction effects when measurements are compared between
frames.\ \ Position-time and momentum-energy are not invariant subspaces
under the action of the group.\ \ The constants $b,c$ and $\hbar
$ are required to describe the conversion of units required in the
mixing rather than just $c$ as in the Poincar\'e case. Again, by
choosing natural units with $c=b=\hbar =1$, these constants are
eliminated. We generally use natural units unless specifically noted.

This is made considerably more subtle than the Poincar\'e case by
the fact that the underlying manifold is now nonabelian. The measurements,
in fact, require the quantum theory involving the Hermitian representation
of the algebra. A frame is now a particle state which is identified
with a state in a unitary irreducible representation of the quaplectic
group. This representation theory is discussed in the section that
follows.

The transformation of the Heisenberg generators between frames is
mathematically the group automorphisms of the algebra. That is,
for $A\in \text{\boldmath $a$}( \mathcal{H}( 1,n) ) $\ \ and\ \ $\Upsilon
\in \mathcal{U}( 1,n) $ with $\Upsilon = e ^{Z}$,\ \ \ the transformed
generators $\tilde{A}$ are 
\begin{equation}
\varsigma ^{\prime }_{\Upsilon }:A\mapsto \tilde{A}=\Upsilon \cdot
A\cdot \Upsilon ^{-1}= e ^{Z}\cdot A\cdot  e ^{-Z}=A+\left[ Z,A\right]
+%
\label{automorphisms of the algebra}
\end{equation}

\noindent With $n=3$, define $J_{i}=\epsilon _{k}^{j,k}L_{j,k}$,
$K_{i}=L_{0,i}$,\ \ $N_{i}=M_{0,i}$\ \ and $R=M_{0,0}$ where $\{L_{a,b},
M_{a,b}\}$ are the generators of $\mathcal{U}( 1,n) $ defined in
$ ($11$ )$.\ \ A general element of the quaplectic algebra is $Z+A$
where
\begin{equation}
\begin{array}{l}
 Z=\beta ^{i}K_{i} +\gamma ^{i}N_{i}+\alpha ^{i} J_{i}+ \theta ^{i,j}
M_{i,j}+\vartheta  R , \\
 A =\frac{t}{\lambda _{t}} T+ \frac{e}{\lambda _{e}}E+\frac{q^{i}}{\lambda
_{q}}Q_{i}+\frac{p^{i}}{\lambda _{p}}P_{i}+ \iota  I .
\end{array}%
\label{Unitary Algebra Element Expanded}
\end{equation}

In this discussion, the Heisenberg algebra degrees of freedom take
on dimensional scales corresponding to the different units of measurement
for time, energy, position,\ \ and momentum. The infinitesimal transformations
$\tilde{A}=A+[Z,A]$ for the basis are
\begin{equation}
\begin{array}{l}
 \tilde{T}=T+\beta ^{i}Q_{i}/c+\gamma ^{i}P_{i}/b+\vartheta  E/c
b, \\
 \tilde{E}=E-c \gamma ^{i}Q_{i}+b \beta ^{i}P_{i}- b c \vartheta
T, \\
 \tilde{Q}_{i}=Q_{i}+\epsilon _{i,j}^{k}\alpha ^{j}Q_{k}+c \beta
^{i}T-\gamma ^{i}E/b+c \theta ^{i,j}P_{j}/b, \\
 \tilde{P}_{i}=P_{i}+\epsilon _{i,j}^{k}\alpha ^{j}Q_{k}+\beta ^{i}E/c+b
\gamma ^{i}T-b \theta ^{i,j}Q_{j}/c.
\end{array}%
\label{canonical transformation equations}
\end{equation}

\noindent And, with $\alpha ^{i}=\theta ^{i,j}=\vartheta =0$, the
{\itshape pure\ \ boost} finite transformations are 
\begin{equation}
\begin{array}{l}
 \tilde{T}= \cosh  \omega  T + \frac{\sinh  \omega }{\omega }\left(
\frac{\beta ^{i}}{c}Q_{i}+ \frac{\gamma ^{i}}{b}P_{i}\right) , \\
 \tilde{E}= \cosh  \omega  E + \frac{\sinh  \omega }{\omega }\left(
-b \gamma ^{i}Q_{i}+c \beta ^{i}P_{i}\right) , \\
 \tilde{Q}_{i}= Q_{i }+ \frac{\cosh  \omega  -1}{\omega ^{2}}\omega
^{i,j}Q_{j} + \frac{\sinh  \omega }{\omega }\left( c \beta ^{i}T-\frac{\gamma
^{i}}{b} E_{i}\right) , \\
 \tilde{P}_{i}= P_{i }+ \frac{ \cosh  \omega  -1}{\omega ^{2}} \omega
^{i,j}P_{j} + \frac{\sinh  \omega }{\omega }\left( b \gamma ^{i}T+\frac{\beta
^{i}}{c}E\right) 
\end{array}%
\label{Finite velocity force boosts}
\end{equation}

\noindent where $\omega ^{i,j}=\beta ^{i}\beta ^{j}+\gamma ^{i}\gamma
^{j}$ and $\omega =\sqrt{\delta _{i,j}\omega ^{i,j}}$ .\ \ \ Note
immediately that if $\gamma ^{i}=0$ these are the equations for
the usual pure Lorentz velocity boost from basic special relativity
for inertial $ ($non-interacting$ )$ frames
\begin{equation}
\begin{array}{l}
 \tilde{T}= \cosh  \beta  T + \frac{\sinh  \beta }{\beta }\frac{\beta
^{i}}{c}Q_{i}, \\
 \tilde{Q}_{i}= Q_{i }+ \frac{ \cosh  \beta  -1}{\beta ^{2}}\beta
^{i}\beta ^{j}Q_{j} + \frac{\sinh  \beta }{\beta }c \beta ^{i}T,
\\
 \tilde{E}= \cosh  \beta  E + \frac{\sinh  \beta }{\beta }c \beta
^{i}P_{i}, \\
 \tilde{P}_{i}= P_{i }+ \frac{ \cosh  \beta  -1}{\beta ^{2}} \beta
^{i}\beta ^{j}P_{j} + \frac{\sinh  \beta }{\beta }\frac{\beta ^{i}}{c}E.
\end{array}%
\label{Finite velocity boosts}
\end{equation}

\noindent Conversely if $\beta ^{i}=0$, these equations reduce to
\begin{equation}
\begin{array}{l}
 \tilde{T}= \cosh  \gamma  T + \frac{\sinh  \gamma }{\gamma }\frac{\gamma
^{i}}{b}P_{i}, \\
 \tilde{P}_{i}= P_{i }+ \frac{ \cosh  \gamma  -1}{\gamma ^{2}}\gamma
^{i}\gamma ^{j}P_{j} + \frac{\sinh  \gamma }{\gamma }b \gamma ^{i}T,
\\
 \tilde{E}= \cosh  \gamma  E + \frac{\sinh  \gamma }{\gamma }b \gamma
^{i}Q_{i}, \\
 \tilde{Q}_{i}= Q_{i }+ \frac{ \cosh  \gamma  -1}{\gamma ^{2}} \gamma
^{i}\gamma ^{j}Q_{j} - \frac{\sinh  \gamma }{\gamma }\frac{\gamma
^{i}}{b}E.
\end{array}%
\label{Finite force boosts}
\end{equation}

\noindent and consequently for this special case, we have a conjugate
set of pure Lorentz boost transformations describing boosts between
frames with relative rate of change of momentum, that is force,
but with negligible velocity. Clearly this an asymptotic case but
it must also be emphasized that a purely, non interacting particle,
is also an asymptotic case, particularly in the Planck regime where
these effects will be manifest. This limit illustrates the bounding
of force that is related to proposed acceleration bounding theories
\cite{cailiello},\cite{Feoli},\cite{Schuller} or minimum length
\cite{Amelino}. These asymptotic cases are just intended to highlight
the properties of the general equations $ ($22$ )$.

The quaplectic group has two different $\mathcal{S}\mathcal{O}(
1,n) \otimes _{s}\mathcal{H}( 1,n) $ subgroups generated by the
sets of generators
\[
\left\{ J_{i},K_{i},E,P_{i},T,Q_{i},I\right\} ,\ \ \ \ \left\{ J_{i},N_{i},E,P_{i},T,Q_{i},I\right\}
\ \ \ \ \ 
\]

\noindent These are the infinitessimal generators that, using by
$ ($20$ )$, may be exponentiated to the finite transformation equations
$ ($24$ )$\ \ and $ ($25$ )$ respectively. $ ($Note in the finite
equations given, the rotation parameters $\alpha ^{i}$ associated
with the generators $J_{i}$ are zero.\ \ \ Each of these in turn
has two distinct Poincar\'e subgroups for the total of four distinct\ \ Poincar\'e
subgroups of the quaplectic group generated by the sets of generators
\[
\begin{array}{ll}
 \left\{ J_{i},K_{i},E,P_{i}\right\} , & \left\{ J_{i},K_{i},T,Q_{i}\right\}
\\
 \left\{ J_{i},N_{i},E,Q_{i}\right\} , & \left\{ J_{i},N_{i},T,P_{i}\right\}
\end{array}
\]

\noindent \noindent The Hermitian representation of the translation
generators of only one of the translation subgroups of these four
Poincar\'e groups can be diagonalized at a time. This may be denoted
by a simple quad, where the representations of the generators on\ \ only
one face may be simultaneously diagonalized. 
\[
\begin{array}{lll}
 \varrho ^{\prime }( T)  & \leftrightarrow  & \varrho ^{\prime }(
Q_{i})  \\
 \updownarrow  &   & \updownarrow  \\
 \varrho ^{\prime }( P_{i})  & \leftrightarrow  & \varrho ^{\prime
}( E) 
\end{array}
\]

Under the special case of pure velocity boosts, this nonabelian
space breaks into invariant abelian time-position and momentum-position
subspaces on which the transformations $ ($24$ )$ act. Likewise,
in the special case of pure force boosts, this nonabelian space
breaks into invariant abelian time-momentum and position-energy
subspaces on which the transformations $ ($25$ )$ act. 

It is important to emphasized that in general, all degrees of freedom
of this non abelian space mix as one transforms from one frame to
another. The physical theory quantum theory that arises from the
unitary representations of the quaplectic group has states for interacting
particles, not just the asymptotic free particle states.\ \ The
unitary representations of the quaplectic group transform one state
into one another and, in so doing, all of the degrees of freedom,
time, position, energy and momentum mix. 

This leads to a a note about the name {\itshape quaplectic} of the
group. The literal meaning of the English word {\itshape qua} is
similar to the preposition {\itshape as}, ' in the role or character
of' and -{\itshape plectic}\ \ has origins in {\itshape pleat} 'to
fold on itself'.\ \ So the literal meaning of {\itshape quaplectic}
is\ \ 'in the character of folding on itself'.\ \ {\itshape Quaplectic}
is also derived from the origins of the group in the {\itshape sym-plectic}
group with the {\itshape qua-ntum}\ \ Heisenberg 'translation' subgroup
and {\itshape qua-d} $ ($as in {\itshape quad} Poincar\'e\ \ subgroups
and the nonabelian {\itshape quad}$ )$ connotations.\ \ 
\section{Unitary representations of the quaplectic group}\label{c:37}\label{Section
unitary reps of quaplectic group}

This section reviews the Mackey theory for unitary irreducible representations
of the\ \ semidirect product group and applies it to the quaplectic
group and algebra. The Heisenberg group is a normal subgroup of
the quaplectic group.\ \ As the\ \ representations of the normal
group are required by the Mackey theory, the representations of
the Heisenberg group are briefly reviewed. As it is, in turn a semidirect
product group, the Mackey theory is again applicable. 
\subsection{Unitary irreducible representations of\ \ semidirect
product groups }

The problem of determining the unitary irreducible representations
of a general class of semidirect product groups has been solved
by Mackey \cite{mackey}. A sufficient condition for the Mackey representation
theory to apply is that the groups are matrix groups that are algebraic.\ \ The
Mackey theorems are reviewed in \cite{Low} and briefly summarized
here. In addition, the manner in which the results lift to the algebra
is given as they are required for the determination of the field
equations. 

Suppose that $\mathcal{A}$ and $ \mathcal{N}$\ \ are matrix groups
that are algebraic with unitary irreducible representations $\xi
$ and $\sigma $ on the respective Hilbert spaces $\text{\boldmath
$\mathrm{H}$}^{\xi }$ and $\text{\boldmath $\mathrm{H}$}^{\sigma
}$. Then for $a\in \mathcal{A}$,\ \ \ and $k\in \mathcal{K}$
\[
\begin{array}{l}
 \xi ( a) :\text{\boldmath $\mathrm{H}$}^{\xi }\rightarrow \text{\boldmath
$\mathrm{H}$}^{\xi }:\left| \phi \right\rangle  \mapsto \left| \tilde{\phi
}\right\rangle  =\xi ( a) \left| \phi \right\rangle  , \\
 \sigma ( k) :\text{\boldmath $\mathrm{H}$}^{\sigma }\rightarrow
\text{\boldmath $\mathrm{H}$}^{\sigma }:\left| \varphi \right\rangle
\mapsto \left| \tilde{\varphi }\right\rangle  =\sigma ( k) \left|
\varphi \right\rangle  .
\end{array}
\]

The general problem is to determine the unitary irreducible representations
$\varrho $, and the Hilbert space $\text{\boldmath $\mathrm{H}$}^{\varrho
}$\ \ on which it acts, of the semidirect product $\mathcal{G}=\mathcal{K}\otimes
_{s}\mathcal{A}$,
 $\varrho ( g) :\text{\boldmath $\mathrm{H}$}^{\varrho }\rightarrow
\text{\boldmath $\mathrm{H}$}^{\varrho }:|\psi \rangle \mapsto |\tilde{\psi
}\rangle =\varrho ( g) |\psi \rangle $.

The Mackey theorems state that these unitary irreducible representations
$\varrho $ may be constructed by first determining the representations\ \ $\varrho
\mbox{}^{\circ}$ of the stabilizer groups, $\mathcal{G}\mbox{}^{\circ}\subseteq
\mathcal{G}$ and then using an induction theorem to obtain the representations
on the full group $\mathcal{G}$. A sufficient condition for the
Mackey group to apply is that $\mathcal{G}$,$ \mathcal{K}$ and $
\mathcal{A}$ are matrix groups that are algebraic in the sense that
they are defined by polynomial constraints on the general linear
groups.

The stabilizer group $\mathcal{G}\mbox{}^{\circ}=\mathcal{K}\mbox{}^{\circ}\otimes
_{s}\mathcal{A}$\ \ where $\mathcal{K}\mbox{}^{\circ}$ is defined
for each of the {\itshape orbits}. These orbits are\ \ defined by
the natural action of elements $k\in \mathcal{K}$ on the unitary
dual $\hat{\mathcal{A}}$ of $\mathcal{A}$. The action defining the
orbits is $k:\hat{\mathcal{A}}\rightarrow \hat{\mathcal{A}}:\xi
\mapsto \tilde{\xi }=k \xi $\ \ where $(k \xi )(a)=\xi ( k\cdot
a\cdot k^{-1}) $ for all $a\in \mathcal{A}$. The little groups $
\mathcal{K}$$ \mbox{}^{\circ}$ are defined by a certain fixed point
condition on each these orbits. 

If $ \mathcal{A}$ is abelian, the fixed point condition is $k \xi
=\xi $ and the representation $\varrho \mbox{}^{\circ}=\sigma \mbox{}^{\circ}\otimes
\chi $ acts on the Hilbert space $\text{\boldmath $\mathrm{H}$}^{\varrho
\mbox{}^{\circ}}\simeq \text{\boldmath $\mathrm{H}$}^{\sigma }\otimes
\mathbb{C}$.\ \ We note that, if $\mathcal{A}$ is abelian, $\mathcal{A}\simeq
\mathbb{R}^{n}$ under addition and the representations are the characters\ \ $\xi
_{c}( a) =\chi _{c}( a) = e ^{i a\cdot c}$ and therefore $\text{\boldmath
$\mathrm{H}$}^{\xi }\simeq \mathbb{C}.$

If $ \mathcal{A}$ is not abelian, the fixed point condition is $k
\xi =\frac{1}{s}\rho ( k) \xi  \rho ( k) ^{-1}$ and the representation
$\varrho \mbox{}^{\circ}=\sigma \mbox{}^{\circ}\otimes \rho $ acts
on the Hilbert space $\text{\boldmath $\mathrm{H}$}^{\varrho \mbox{}^{\circ}}\simeq
\text{\boldmath $\mathrm{H}$}^{\sigma \mbox{}^{\circ}}\otimes \text{\boldmath
$\mathrm{H}$}^{\xi }$.\ \ $\rho $ is a projective extension of the
representation $\xi $ to\ \ $\mathcal{G}\mbox{}^{\circ}$, $\rho
( g) :\text{\boldmath $\mathrm{H}$}^{\xi }\rightarrow \text{\boldmath
$\mathrm{H}$}^{\xi }$\ \ for $g\in \mathcal{G}\mbox{}^{\circ}$ with
$\rho |_{\mathcal{A}}\simeq \xi $. If $\mathcal{A}$ is abelian,
the extension is trivial, $\rho |_{\mathcal{K}}\simeq 1$ and this
reduces to the abelian case above. 

These representations may be lifted to the algebra. Define $T_{e}\xi
=\xi ^{\prime }$,\ \ $T_{e}\sigma \mbox{}^{\circ}=\sigma \mbox{}^{\circ \prime
}$\ \ and $T_{e}\varrho \mbox{}^{\circ}=\varrho \mbox{}^{\circ \prime
}$. Then for $A\in \text{\boldmath $\mathrm{a}$}( \mathcal{A}) \simeq
T_{e}\mathcal{A}$, $Z\in \text{\boldmath $\mathrm{a}$}( \mathcal{K}\mbox{}^{\circ})
$ and $W=A+Z\in \text{\boldmath $\mathrm{a}$}( \mathcal{G}\mbox{}^{\circ})
$ we have
\begin{equation}
\begin{array}{l}
 \varrho \mbox{}^{\circ \prime }( W) :\text{\boldmath $\mathrm{H}$}^{\varrho
\mbox{}^{\circ}}\rightarrow \text{\boldmath $\mathrm{H}$}^{\varrho
\mbox{}^{\circ}} \\
 =\sigma \mbox{}^{\circ \prime }( Z) \oplus \rho ^{\prime }( W)
:\text{\boldmath $\mathrm{H}$}^{\sigma \mbox{}^{\circ}}\otimes \text{\boldmath
$\mathrm{H}$}^{\xi }\rightarrow \text{\boldmath $\mathrm{H}$}^{\sigma
\mbox{}^{\circ}}\otimes \text{\boldmath $\mathrm{H}$}^{\xi } \\
 :\left| \psi \right\rangle  \mapsto \left| \tilde{\psi }\right\rangle
=\sigma \mbox{}^{\circ \prime }( Z) \left| \varphi \right\rangle
\otimes \left| \phi \right\rangle  \oplus \left| \varphi \right\rangle
\otimes \rho ^{\prime }( W) \left| \phi \right\rangle  .
\end{array}
\end{equation}

The basis of the algebra satisfies the Lie algebra 
\begin{equation}
\begin{array}{l}
 \left[ A_{\mu },A_{\nu }\right] =c_{\mu ,\nu }^{\lambda } A_{\lambda
}, \\
 \left[ Z_{\alpha },Z_{\beta }\right] =c_{\alpha ,\beta }^{\gamma
} Z_{\gamma }, \\
 \left[ A_{\mu },Z_{\alpha }\right] =c_{\mu ,\alpha }^{\nu } A_{\nu
}.
\end{array}
\end{equation}

\noindent where $\alpha ,\beta ..=1...\dim ( \mathcal{K}) $ and
$\mu ,\nu ..=1,..\dim ( \mathcal{A}) $. Then, the Hermitian projective
extension representation $\rho$ of the generators satisfies the
commutation relations 
\begin{equation}
\begin{array}{l}
 \left[ \rho ^{\prime }( A_{\mu }) ,\rho ^{\prime }( A_{\nu }) \right]
=i c_{\mu ,\nu }^{\lambda } \rho ^{\prime }( A_{\lambda }) , \\
 \left[ \rho ^{\prime }( Z_{\alpha }) ,\rho ^{\prime }( Z_{\beta
}) \right] =i \frac{1}{s}c_{\alpha ,\beta }^{\gamma } \rho ^{\prime
}( Z_{\gamma }) , \\
 \left[ \rho ^{\prime }( A_{\mu }) ,\rho ^{\prime }( Z_{\alpha })
\right] =i \frac{1}{s}c_{\mu ,\alpha }^{\nu } \rho ^{\prime }( A_{\nu
}) .
\end{array}%
\label{projective rep commutators}
\end{equation}

As the $\rho ^{\prime }( Z_{\alpha }) $ act on the Hilbert space
$\text{\boldmath $\mathrm{H}$}^{\xi }$, they must be elements of
the enveloping algebra $\text{\boldmath $\mathrm{a}$}( \mathcal{A})
\simeq \text{\boldmath $\mathrm{a}$}( \mathcal{A}) \oplus \text{\boldmath
$\mathrm{a}$}( \mathcal{A}) \otimes \text{\boldmath $\mathrm{a}$}(
\mathcal{A}) \oplus ...$ .\ \ and therefore 
\begin{equation}
\varrho  ^{\prime }\left( Z_{\alpha }\right) = d_{\alpha }^{\mu
}\xi  ^{\prime }\left( A_{\mu }\right) + d_{\alpha }^{\mu ,\nu }\xi
^{\prime }\left( A_{\mu }\right) \xi  ^{\prime }\left( A_{\nu }\right)
+....%
\label{rho enveloping in terms of xi}
\end{equation}

\noindent These may be substituted into the commutation relations
above to determine the constants $\{d_{\alpha }^{\mu },d_{\alpha
}^{\mu ,\nu },...\}$.

We have now characterized the representations $\varrho \mbox{}^{\circ}$
acting on $\text{\boldmath $\mathrm{H}$}^{\varrho \mbox{}^{\circ}}$
and it remains to use the Mackey induction theorem to obtain the
representations on the full group $\mathcal{G}$.\ \ Clearly, if
$\mathcal{G}\mbox{}^{\circ}\simeq \mathcal{G}$, this implies that
$\varrho =\varrho \mbox{}^{\circ}$ and $\text{\boldmath $\mathrm{H}$}^{\varrho
\mbox{}^{\circ}}=\text{\boldmath $\mathrm{H}$}^{\varrho } $ and
the induction is trivial and the representation $\varrho $ of $\mathcal{G}$
is determined.

However, if $\mathcal{G}\mbox{}^{\circ}$ is a proper closed subgroup
of $\mathcal{G}$, then the Mackey induction theorem states that
the representations act on the Hilbert space $\text{\boldmath $\mathrm{H}$}^{\varrho
}\simeq \text{\boldmath $\mathrm{H}$}^{\sigma \mbox{}^{\circ}}\otimes
\text{\boldmath $\mathrm{L}$}^{2}( \mathbb{A},\text{\boldmath $\mathrm{H}$}^{\xi
},\mu ) $ with $\mathbb{A}\simeq \mathcal{K}/\mathcal{K}\mbox{}^{\circ}$\ \ where
for $|\phi \rangle \in \text{\boldmath $\mathrm{L}$}^{2}( \mathbb{A},\text{\boldmath
$\mathrm{H}$}^{\xi },\mu ) $, then\ \ $\phi :\mathbb{A}\rightarrow
\text{\boldmath $\mathrm{H}$}^{\xi }:a\mapsto |\phi _{a}\rangle
=\langle a|\phi \rangle $. Also, if $\mathcal{A}$ is abelian, $\text{\boldmath
$\mathrm{H}$}^{\xi }\simeq \mathbb{C}$ and this reduces to $\phi
\in \text{\boldmath $\mathrm{L}$}^{2}( \mathbb{A},\mathbb{C},\mu
) $ with $\phi :\mathbb{A}\rightarrow \mathbb{C}:a\mapsto \phi (
a) $.

Thus if $|\tilde{\psi }\rangle =\varrho ( g) |\psi \rangle $ for
$g\in \mathcal{G}$ and $|\psi \rangle ,|\tilde{\psi }\rangle \in
\text{\boldmath $\mathrm{H}$}^{\varrho }$,\ \ then $|\tilde{\psi
}_{\tilde{a}}\rangle =\langle \tilde{a}|\tilde{\psi }\rangle =\langle
\tilde{a}|\varrho ( g) |\psi \rangle $\ \ where $|\tilde{\psi }_{\tilde{a}}\rangle
\in \text{\boldmath $\mathrm{H}$}^{\varrho \mbox{}^{\circ}}$.\ \ The
induction theorem then states that
\[
\left| \tilde{\psi }_{\tilde{a}}\right\rangle  = \varrho \mbox{}^{\circ}(
\Theta ( \tilde{a}) ^{-1} g \Theta ( a) )  \left| \psi _{a}\right\rangle
\mathrm{with}\ \ \ \tilde{a}=g a.
\]

\noindent where $\Theta $ is the natural section $\Theta :\mathbb{A}\rightarrow
\mathcal{G}:a\mapsto g=\Theta ( a) $.\ \ \ Note that for $\mathit{g}\mbox{}^{\circ}\in
\mathcal{G}\mbox{}^{\circ}$, $\mathit{g}\mbox{}^{\circ} a=a=\tilde{a}$
and as $\Theta ( a) \in \mathcal{G}\mbox{}^{\circ}$, this reduces
to just an inner automorphism of\ \ $\mathcal{G}\mbox{}^{\circ}$
that defines the equivalence classes of\ \ $ \rho \mbox{}^{\circ}$
and this reduces to the expected\ \ $|\tilde{\psi }_{a}\rangle =
\varrho \mbox{}^{\circ}( \mathit{g}\mbox{}^{\circ})  |\psi _{a}\rangle
$. Putting it together, we have the induced representation theorem
where the representations $\varrho \mbox{}^{\circ}$ of $ \mathcal{G}$$
\mbox{}^{\circ}$ on $\text{\boldmath $\mathrm{H}$}^{\varrho \mbox{}^{\circ}}$
are induced onto the representations $\varrho $ of $ \mathcal{G}$
on\ \ $\text{\boldmath $\mathrm{H}$}^{\varrho }$ by
\[
\left\langle  a\right| \varrho ( g)  \left| \psi \right\rangle 
=\left| \tilde{\psi }_{a}\right\rangle  =\varrho \mbox{}^{\circ}(
\Theta ( a) ^{-1}\cdot  g\cdot  \Theta ( g^{-1}a) ) \left| \psi
_{g^{-1}a}\right\rangle  .
\]

Again, in the abelian case, $|\psi _{g^{-1}a}\rangle \in \text{\boldmath
$\mathrm{H}$}^{\xi }\simeq \mathbb{C}$\ \ in which case this is
written simply as\ \ \ $\psi ( g^{-1}a) $
\subsection{Mackey representations of the Heisenberg group}\label{c:26}

The Heisenberg group $\mathcal{H}( n+1) \simeq \mathcal{T}( n+1)
\otimes _{s}\mathcal{T}( n+2) $\ \ is the normal subgroup of the
quaplectic group and we therefore review the Mackey representation
theory \cite{Low},\cite{Major} of the Heisenberg group briefly as
it is required for the representations of the quaplectic\ \ group.\ \ 

$\mathcal{A}\simeq \mathcal{T}( n+2) $ is the normal subgroup with
an algebra spanned by $\{I,Y_{a}\}$ and $\mathcal{K}\simeq \mathcal{T}(
n+1) $ is the\ \ homogeneous group with an algebra spanned by\ \ $\{X_{a}\}$.\ \ We
use $\chi $ for the notation for the representation of\ \ the normal
group $\mathcal{A}$ in this section which, for the translation group,
are just the characters
\begin{equation}
\begin{array}{rl}
 \chi _{c,u}( h( 0,y,\iota ) ) \left| z\right\rangle   & = e ^{i
\left( \iota  \chi ^{\prime }( I)  +y^{a}\chi ^{\prime }( Y_{a})
\right) }\left| z\right\rangle   \\
  & = e ^{i \left( \iota  c+y\cdot  u\right) }\left| z\right\rangle
,
\end{array}%
\label{unitary rep of translation group as Heisenberg normal subgroup}
\end{equation}

\noindent where $\chi ^{\prime }( I) |z\rangle =c|z\rangle $ and
$\chi ^{\prime }( Y_{a}) |z\rangle =u_{a}|z\rangle $. $c\in \mathbb{R}$,
$u\in \mathbb{R}^{n+1}$ are the Casimir eigenvalues of the translation
group and $|z\rangle \in \text{\boldmath $\mathrm{H}$}^{\chi }\simeq
\mathbb{C}$.\ \ The action on the dual $\hat{\mathcal{A}}$ defining
the orbits is\ \ \ 
\[
\begin{array}{l}
 \left( h( x,0,0) \chi _{c,u} \right) \left( h( 0,y,\iota ) \right)
\\
 =\chi _{u,c}\ \ \left( h( x,0,0) \cdot h( 0,x,\iota ) \cdot h(
x,0,0) ^{-1}\right)  \\
 =\chi _{u+c y,c} \left( h( 0,y,\iota ) \right) ,
\end{array}
\]

\noindent for all $h( 0,y,\iota ) \in \mathcal{T}( n+1) $.\ \ Therefore,
the fixed point condition is $\chi _{u,c} =\chi _{u+c y,c}$.\ \ One
solution is $c=0$ in which case the representation $\chi ^{\prime
}( I) |z\rangle =0$ and hence the representation is a degenerative
case equivalent to the representations of $\mathcal{T}( 2n) $ that
are not considered further.\ \ For $c\neq 0$, the fixed point is
satisfied only if $u=0$ and so the little group is trivial, $\mathcal{K}\mbox{}^{\circ}\simeq
e$. Consequently, $\varrho \mbox{}^{\circ}=\chi $. The homogeneous
space is $\mathbb{A}\simeq \mathcal{K}/\mathcal{K}\mbox{}^{\circ}\simeq
\mathcal{T}( n+1) /e \simeq \mathbb{R}^{n+1}$.\ \ The Hilbert space
is therefore
\begin{equation}
\text{}\text{\boldmath $\mathrm{H}$}^{\varrho }\simeq \text{\boldmath
$\mathrm{L}$}^{2}( \mathbb{A},\text{\boldmath $\mathrm{H}$}^{\xi
},\mu ) \simeq \text{\boldmath $\mathrm{L}$}^{2}( \mathbb{R}^{n+1},\mathbb{C},\mu
) %
\label{Heisenberg Hilbert space}
\end{equation}

\noindent as expected. $I$ is the only Casimir invariant operator
and its eigenvalues label the representations.\ \ Points $a_{x}\in
\mathbb{A}$ are labeled by $h( x,0,0) \in \mathcal{K}$ with $\Theta
( a_{x}) =h( x,0,0) $.\ \ $h( x,y,\iota ) a_{\tilde{x}}=a_{\tilde{x}+x}$
and therefore the Mackey induction theorem yields
\[
\begin{array}{rl}
 \tilde{\psi }( a_{y})  & =\left\langle  a_{x}\right| \varrho (
h( \tilde{x},\tilde{y},\iota ) )  \left| \psi \right\rangle   \\
  & =\varrho \mbox{}^{\circ}( h( 0,\tilde{y},\iota -\tilde{y}\cdot
x ) ) \psi ( a_{x-\tilde{x}}) .
\end{array}
\]

\noindent where a calculation showing that\ \ \ 
\[
h( 0,\tilde{y},\iota -\tilde{y}\cdot x ) =h( \tilde{x},\tilde{y},\iota
) \cdot  \Theta ( h( \tilde{x},\tilde{y},\iota ) ^{-1}a_{x}) 
\]

\noindent and $a_{x-\tilde{x}}=h(\tilde{x},\tilde{y},\iota )^{-1}a_{x}$
has been used.\ \ Using\ \ the bijection $a_{x}\leftrightarrow x$
and the definition of\ \ $\varrho \mbox{}^{\circ}=\chi $ in $ ($30$
)$ with $u=0$\ \ this gives 
\begin{equation}
\tilde{\psi }( x) = e ^{i \left( \iota -\tilde{y}\cdot x\right)
c}\psi ( x-\tilde{x}) =  e ^{i \iota  c} e ^{-i c\tilde{y}\cdot
x} e ^{i\ \ \tilde{x} \cdot i \frac{\partial }{\partial x}}\psi
( x) .%
\label{heisenberg unitary rep}
\end{equation}

It follows that the representation of the algebra is
\begin{equation}
\begin{array}{l}
 \left\langle  x\right| \varrho ^{\prime }( I) \left| \psi \right\rangle
=c \left| \psi \right\rangle   , \\
 \left\langle  x\right| \varrho ^{\prime }( X_{a}) \left| \psi \right\rangle
=c\ \ x^{a}\left| \psi \right\rangle   , \\
  \left\langle  x\right| \varrho ^{\prime }( Y_{a}) \left| \psi
\right\rangle  =i \frac{\partial }{\partial  x}\left| \psi \right\rangle
.
\end{array}
\end{equation}

Thus, it is clear that the usual basic representations of quantum
mechanical position and momentum Hermitian operators and their associated
Hilbert space are directly computed using the Mackey representation
theory of the Heisenberg group.
\subsection{Mackey representations of the quaplectic group}\label{c:26}

The normal subgroup of the quaplectic group is the Heisenberg group
and so the Mackey case for nonabelian normal subgroups applies.
As the Heisenberg is now the normal subgroup, we denote the representations
$\varrho $ from the previous section given in $ ($32$ )$ in this
section by $\xi $. These representations $\xi _{c}$ of $\mathcal{H}(
n+1) $ are labeled by the eigenvalues $c$ of the Casimir invariant
operator $I$.\ \ \ The case $c=0$ corresponds to the degenerate
abelian case, that we do not consider further here.

The general nonabelian case corresponds to $c\in \mathbb{R}\backslash
\{0\}$.\ \ This requires the definition of a projective extension
$\rho $ of\ \ $\xi $ with the property that $\rho |_{\mathcal{H}(
n+1) }=\xi $.\ \ The action of $\Upsilon \in \mathcal{U}( 1,n) $
on the representations $\xi  $ of the normal subgroup is
\[
\begin{array}{l}
 \left( \Upsilon  \xi _{c}\right) \left( g( 0,z,\iota ) \right)
=\left( \Upsilon  \rho \right) \left( g( 0,z,\iota ) \right)   \\
 =\rho ( g( \Upsilon ,0,0) \cdot g( 0,z,\iota ) \cdot g( \Upsilon
,0,0) ^{-1})  \\
 =\rho ( \Upsilon ) \cdot \xi _{c}( g( 0,z,\iota ) ) \cdot \rho
( \Upsilon ^{-1}) 
\end{array}
\]

\noindent for all $g( 0,z,\iota ) \in \mathcal{H}( n+1) $.\ \ \ Consequently,
the fixed point condition for the nonabelian case\ \ 
\[
\Upsilon \cdot \xi _{c}=\rho ( \Upsilon ) \cdot \xi _{c}\cdot \rho
( \Upsilon ) ^{-1}
\]

\noindent is identically satisfied for all $\Upsilon \in \mathcal{U}(
1,n) $. Consequently, the little group is\ \ $\mathcal{K}\mbox{}^{\circ}\simeq
\mathcal{U}( 1,n) \simeq \mathcal{K}$ \cite{Wolf}.

An immediate consequence of this is that the Mackey induction is
trivial and the Hilbert space of the representations is given simply
by the direct product of the Hilbert space of the representations
of the little group and the Hilbert space of the representations
of the Heisenberg group, $\text{\boldmath $\mathrm{H}$}^{\varrho
}=\text{\boldmath $\mathrm{H}$}^{\sigma }\otimes \text{\boldmath
$\mathrm{H}$}^{\xi } $.\ \ As the Hilbert space $\text{\boldmath
$\mathrm{H}$}^{\xi }$ of the representations $\xi $ of $\mathcal{H}(
1,n) $ have been determined in $ ($31$ )$, this is 
\[
\text{\boldmath $\mathrm{H}$}^{\varrho }=\text{\boldmath $\mathrm{H}$}^{\sigma
}\otimes \text{\boldmath $\mathrm{L}$}^{2}( \mathbb{R}^{n+1},\mathbb{C},\mu
) .
\]

$\text{\boldmath $\mathrm{H}$}^{\sigma }$ is generally a countably
infinite vector space as the unitary irreducible representations
of $\mathcal{U}( 1,n) $ are generally infinite dimensional. This
is remarkably different from the Poincar\'e case where the Hilbert
space was over the $n$ dimensional null and time-like hyperboloid
surfaces in $\mathbb{R}^{n+1}$. The complete set of irreducible
representations in that case is required to foliate $\mathbb{R}^{n+1}$\ \ Here\ \ a
single representation covers the entire space $\mathbb{R}^{n+1}$
and so the particle states are not constrained by the representation
to a hypersurface. 

The representations $\varrho $ are the direct product of $\sigma
$ and $\rho$ given by
\[
\begin{array}{l}
 \varrho ( g( \Upsilon ,z,\iota ) ) :\text{\boldmath $\mathrm{H}$}^{\varrho
}\rightarrow \text{\boldmath $\mathrm{H}$}^{\varrho  }:\left| \psi
\right\rangle  \simeq \left| \varphi \right\rangle  \otimes \left|
\phi \right\rangle  \mapsto \left| \tilde{\psi }\right\rangle  
\\
 \left. \left. \left. =\varrho ( g( \Upsilon ,z,\iota ) ) \left|
\psi \right. \right\rangle  \simeq \sigma ( g( \Upsilon ,0,0) )
\left| \varphi \right. \right\rangle  \otimes \rho ( g( \Upsilon
,z,\iota ) ) \left| \phi \right. \right\rangle  .
\end{array}
\]

\noindent where, as above,\ \ $|\psi \rangle \in \text{\boldmath
$\mathrm{H}$}^{\varrho }$, $ |\varphi \rangle \in \text{\boldmath
$\mathrm{H}$}^{\sigma }$,\ \ $|\phi \rangle \in \text{\boldmath
$\mathrm{H}$}^{\xi }\simeq \text{\boldmath $\mathrm{L}$}^{2}( \mathbb{R}^{n+1},\mathbb{C},\mu
) $.\ \ 

The full representations $\varrho $ may be lifted to the algebra
$\varrho ^{\prime }$ to act on a basis $\{A_{b}^{\pm },Z_{a,b}\}$
of $\text{\boldmath $\mathrm{a}$}( \mathcal{Q}( 1,n) ) $.\ \ Using
$ ($12$ )$ and noting that as the representation is Hermitian, the
appropriate factors of $i$ must be inserted, 
\begin{equation}
\begin{array}{l}
 \left[ \varrho ^{\prime }( A_{a}^{+}) ,\varrho ^{\prime }( A_{b}^{-})
\right] = \eta _{a,b}\varrho ^{\prime }( \mathit{I}\mathit{)} ,
\\
 \left[ \varrho ^{\prime }( Z_{a,b}) ,\varrho ^{\prime }( A_{c}^{\pm
}) \right] =\mp \eta _{a,c}\varrho ^{\prime }( A_{b}^{\pm } ) ,
\\
 \left[  \varrho ^{\prime }( Z_{a,b}) ,\varrho ^{\prime }( Z_{c,d})
\right]  =\varrho ^{\prime }(  Z_{c,b})  \eta _{a,d}-\varrho ^{\prime
}( Z_{a,d})  \eta _{b,c}.
\end{array}%
\label{Representation of Canonical Algebra}
\end{equation}

The representations $\rho $ may also be lifted to the algebra $\rho
^{\prime }$ to act on a basis $Z_{a,b}$ of $\text{\boldmath $\mathrm{a}$}(
\mathcal{U}( 1,n) ) $.\ \ Using\ \ $ ($28$ )$, the representations
of the algebra are
\[
\begin{array}{l}
 \left[ \xi ^{\prime }( A_{a}^{-}) ,\xi ^{\prime }( A_{b}^{+}) \right]
= \eta _{a,b} \xi ^{\prime }( I) ,\ \  \\
 \left[ \rho ^{\prime }( Z_{a,b}) ,\xi ^{\prime }( A_{c}^{\pm })
\right] =\pm \frac{1}{s}\eta _{a,c}\xi ^{\prime }( A_{b}^{\pm })
, \\
 \left[ \rho ^{\prime }( Z_{a,b}) ,\rho ^{\prime }( Z_{a,b}) \right]
=\frac{1}{s}\left( \eta _{b,c}\rho ^{\prime }( Z_{a,d}) -\eta _{a,d}\rho
^{\prime }( Z_{c,b}) \right)  .
\end{array}
\]

\noindent These elements must be in the enveloping algebra $\text{\boldmath
$\mathrm{e}$}( \mathcal{U}( 1,n) ) $ and therefore may be expressed,
using $ ($29$ )$, as
\[
\rho ^{\prime }( Z_{a,b}) =d_{\pm }^{a}\xi ^{\prime }( A_{a}^{\pm
}) + d_{\pm ,\pm }^{a,b}\xi ^{\prime }( A_{a}^{\pm }) \xi ^{\prime
}( A_{b}^{\pm }) +...
\]

\noindent Substituting into the above commutation relations gives\ \ 
\begin{equation}
\rho ^{\prime }( Z_{a,b}) =\frac{1}{s}\xi ^{\prime }( A_{a}^{+})
\xi ^{\prime }( A_{b}^{-}) .%
\label{rho Z in terms of xi A}
\end{equation}

\noindent The representation of the algebra is given by
\[
\begin{array}{l}
 \varrho ^{\prime }( I) \left| \psi \right\rangle  =\left| \varphi
\right\rangle  \otimes \xi ^{\prime }( I) \left| \phi \right\rangle
, \\
 \varrho ^{\prime }( A_{a}^{\pm }) \left| \psi \right\rangle  =\left|
\varphi \right\rangle  \otimes \xi ^{\prime }( A_{a}^{\pm }) \left|
\phi \right\rangle  ,\ \  \\
 \varrho ^{\prime }( Z_{a,b}) \left| \psi \right\rangle  =\sigma
^{\prime }( Z_{a,b}) \left| \varphi \right\rangle  \otimes \left|
\phi \right\rangle  \oplus \left| \varphi \right\rangle  \otimes
\rho ^{\prime }( Z_{a,b}) \left| \phi \right\rangle  
\end{array}
\]

\noindent where $\rho ^{\prime }( Z_{a,b}) $ is defined in $ ($36$
)$. The matrix elements of $\xi ^{\prime }( A_{a}^{\pm }) $ and
$\rho ^{\prime }( Z_{a,b}) $ with respect to a basis $|x\rangle
$ are given by 
\begin{equation}
\begin{array}{l}
 \left\langle  x\right| \xi ^{\prime }( A_{a}^{\pm }) \left| \psi
\right\rangle   =\frac{1}{\sqrt{2}} \left( x\pm \frac{\partial }{\partial
x^{a}}\right)  , \\
 \left\langle  x\right| \rho ^{\prime }( Z_{a,b}) \left| \psi \right\rangle
=\frac{1}{2} \text{$ (x^{a}+\frac{\partial }{\partial x^{a}}$$ )$$
(x^{a}-\frac{\partial }{\partial x^{a}})$}.
\end{array}%
\label{xi rep of A}
\end{equation}

\noindent The matrix elements for the $\sigma ^{\prime }( Z_{a,b})
$ are countably infinite matrices that will be defined further shortly,
\[
\left( \Sigma _{a,b}\right) _{\tilde{M},M}=\left\langle  \tilde{M}\right|
\sigma ^{\prime }( Z_{a,b}) \left| M\right\rangle  
\]

\noindent \noindent where $\tilde{M},M$ take values in a countably
infinite set.\ \ \ Then, the expression for the unitary irreducible
representations of the group may be written explicitly as
\[
\begin{array}{l}
 \left\langle  \tilde{M},x\right| \varrho ( g( \Upsilon ( \zeta
) ,z,\iota ) ) \left| \psi \right\rangle  = \\
  e ^{i \zeta ^{a,b}( \Sigma _{a,b}) _{\tilde{M},M}}\otimes  e ^{i(
\iota I+\frac{1}{\sqrt{2}}z_{\pm }^{a} \left( x^{a}\pm \frac{\partial
}{\partial x^{a}}\right) ) }\times  \\
  e ^{i\frac{1}{2} \zeta ^{a,b} \left( x^{a}+\frac{\partial }{\partial
x^{a}}\right) \left( x^{b}-\frac{\partial }{\partial x^{b}}\right)
}\psi _{M}( x) 
\end{array}
\]

\noindent with $z_{+}^{a}=z^{a}$ and $z_{-}^{a}=\overline{z}^{a}$.\ \ This
completes the characterization of the unitary irreducible representations
of the quaplectic group and algebra given by the Mackey theory.\ \ 

We now characterize the representations of the algebra and the Hilbert
spaces on which they act in further detail. Following this, the
form of the general field equations for the quaplectic group are
given and specific cases studied. 
\subsection{The representations $\rho ^{\prime }$\ \ of the algebra
of $\mathcal{Q}( 1,n) $.}\label{c:38}

The representations $\rho ^{\prime }$ of the algebra of the full
group $\mathcal{Q}( 1,n) $ act on the Hilbert space\ \ $\text{\boldmath
$\mathrm{H}$}^{\xi }\simeq \text{\boldmath $\mathrm{L}$}^{2}( \mathbb{R}^{n+1},\mathbb{C})
$ of the representations $\xi $ of the normal subgroup $\mathcal{H}(
1,n) $.\ \ \ The representations $\rho ^{\prime }$ are the extension
to the full set of generators $\{Z_{a,b},A_{a}^{\pm },I\}$, $a,b,...=0,1,...n$
of\ \ $\mathcal{Q}( 1,n) $ where $\rho ^{\prime }( A_{a}^{\pm })
=\xi ^{\prime }( A_{a}^{\pm }) $ and $\rho ^{\prime }( I) =\xi ^{\prime
}( I) $ and $\rho ^{\prime }( Z_{a,b}) $\ \ is given by equation
$ ($35$ )$. 

A coherent basis for the Hilbert space $\text{\boldmath $\mathrm{H}$}^{\xi
}$ may be defined by the complete set of orthonormal states $|\eta
_{k_{0},k_{1},...k_{n}}\rangle $ \cite{Kim}
\begin{equation}
\left\langle  x_{a}|\eta _{k_{a}}\right\rangle  =\eta _{k_{a}}(
x_{a})  =\frac{1 }{\sqrt{\pi }2^{k_{a}}k_{a}!} e ^{-\frac{1}{2}\left(
x_{a}\right) ^{2}}H_{k_{a}}( x_{a}) ,%
\label{heisenberg hermite polynomial}
\end{equation}

\noindent with $k_{a}\in \mathbb{Z}^{+}$, $x_{a}\in \mathbb{R}$
and\ \ $|\eta _{K}\rangle  =|\eta _{k_{0,}k_{1},...k_{n}}\rangle
$ with $K=(k_{0}, k_{1},....k_{n})$.

\noindent Defining $\text{$I_{a}$}$ now to be an $n+1$ tuple of
0's with a 1 at the $a \mathrm{th}$ position, $\text{$I_{a}$}=(0,0,,,1,...0)$
, the action of the representation on the basis is
\begin{equation}
\begin{array}{l}
 \xi ^{\prime }( A_{0}^{-}) \left| \eta _{K}\right\rangle   =\sqrt{k_{0}+1}\left|
\eta _{K+I_{0}}\right\rangle  , \\
 \xi ^{\prime }( A_{0}^{+}) \left| \eta _{K}\right\rangle   =\sqrt{k_{0}}\left|
\eta _{K-I_{0}}\right\rangle  , \\
 \xi ^{\prime }( A_{i}^{+}) \left| \eta _{K}\right\rangle   =\sqrt{k_{i}+1}\left|
\eta _{K+I_{i}}\right\rangle  , \\
 \xi ^{\prime }( A_{i}^{-}) \left| \eta _{K}\right\rangle   =\sqrt{k_{i}}\left|
\eta _{K-I_{i}}\right\rangle   ,
\end{array}%
\label{coherent basis A}
\end{equation}

\noindent where as always $i,j=1,2,..n$.\ \ It follows directly
that 
\begin{equation}
\begin{array}{l}
 \rho ^{\prime }( Z_{i,0}) \left| \eta _{K}\right\rangle   =\frac{1}{s}\sqrt{\left(
k_{0}+1\right) \left( k_{i}+1\right) }\left| \eta _{K+I_{i}+I_{0}}\right\rangle
, \\
 \rho ^{\prime }( Z_{0,i}) \left| \eta _{K}\right\rangle   =\frac{1}{s}\sqrt{k_{i}k_{0}}\left|
\eta _{K-I_{i}-I_{0}}\right\rangle  , \\
 \rho ^{\prime }( Z_{j,i}) \left| \eta _{K}\right\rangle   =\frac{1}{s}\sqrt{\left(
k_{j}+1\right) k_{i}}\left| \eta _{K-I_{i}+I_{j}}\right\rangle 
, i\neq j, \\
 \rho ^{\prime }( Z_{a,a}) \left| \eta _{K}\right\rangle   =\frac{1}{s}
k_{a}\left| \eta _{K}\right\rangle  .
\end{array}%
\label{coherent basis Z}
\end{equation}

The little group may be factored as $\mathcal{U}( 1,n) =\mathcal{U}(
1) \otimes \mathcal{S}\mathcal{U}( 1,n) .\text{}$ Note that the
generator $U$ of the algebra of $\mathcal{U}( 1) $ is defined by
$U=\eta ^{a,b}Z_{a,b}$ and so the representation is\ \ 
\begin{equation}
\rho ^{\prime }( U) \left| \eta _{K}\right\rangle   =\left( -k_{0
}+k_{1}+k_{2}+...k_{n}\right) \left| \eta _{K}\right\rangle  =k\left|
\eta _{K}\right\rangle   %
\label{k definition}
\end{equation}

\noindent with $k=-k_{0 }+\sum \limits_{i=1}^{n}k_{i}$ where again
$k_{a}\in \mathbb{Z}\geq 0$.\ \ The Hilbert space $\text{\boldmath
$\mathrm{H}$}^{\xi }$\ \ may be written as a direct sum $\text{\boldmath
$\mathrm{H}$}^{\xi } = \bigoplus \limits_{k=1}^{\infty }\text{\boldmath
$\mathrm{H}$}_{k}^{\xi }$ of subspaces $\text{\boldmath $\mathrm{H}$}_{k}^{\xi
}$ that are invariant under the $\rho ^{\prime }$representations\ \ of
the generators $Z_{a,b}$ of $\text{\boldmath $a$}( \mathcal{U}(
1,n) ) $ 
\begin{equation}
\rho ^{\prime }( Z_{a,b}) :\text{\boldmath $\mathrm{H}$}_{k}^{\xi
}\rightarrow \text{\boldmath $\mathrm{H}$}_{k}^{\xi }:\left| \tilde{\eta
}_{K}\right\rangle  \mapsto \rho ^{\prime }( Z_{a,b}) \left| \eta
_{K}\right\rangle  .%
\label{Z invariant subspace}
\end{equation}

\noindent The representations of the generators $\rho ^{\prime }(
A_{a}^{\pm }) $\ \ of $\text{\boldmath $a$}( \mathcal{H}( 1,n) )
$ cause transitions between $\text{\boldmath $\mathrm{H}$}_{k}^{\xi
}$ subspaces with different $ k$. 
\[
\rho ^{\prime }( A_{i}^{\pm }) :\text{\boldmath $\mathrm{H}$}_{k}^{\xi
}\rightarrow \text{\boldmath $\mathrm{H}$}_{k\pm 1}^{\xi },\ \ \rho
^{\prime }( A_{0}^{\pm }) :\text{\boldmath $\mathrm{H}$}_{k}^{\xi
}\rightarrow \text{\boldmath $\mathrm{H}$}_{k\mp 1}^{\xi }
\]

\noindent As $ \mathcal{U}( 1,n) $ is not compact, each $\text{\boldmath
$\mathrm{H}$}_{k}^{\xi }$ is countably infinite dimensional due
to the indefinite signature.\ \ Define\ \ \ \ $l=\sum \limits_{i=1}^{n}k_{i}$,
and the $l$ label the finite dimensional $\mathcal{U}( n) $ invariant
subspaces $\text{\boldmath $\mathrm{H}$}_{l}^{\xi }\simeq \text{\boldmath
$\mathrm{V}$}^{l}$.\ \ Thus, the $\text{\boldmath $\mathrm{H}$}_{k}^{\xi
}$ may, in turn, be written as an infinite direct sum of the finite
dimensional Hilbert spaces\ \ $\text{\boldmath $\mathrm{H}$}_{k}^{\xi
}=\bigoplus \limits_{l=k}^{\infty }\text{\boldmath $\mathrm{V}$}^{l}$.
\subsection{The representations $\sigma ^{\prime }$ of the algebra
of $\mathcal{U}( 1,n) $.}\label{c:39}

The next task is to give the explicit form of the representations
$\sigma ^{\prime }( Z_{a,b}) $ of the algebra of $\mathcal{U}( 1,n)
$.\ \ This problem has been solved both for the representation of
the algebra and also for the group in \cite{Maekawa},\cite{Ottoson},\cite{Gitman}.\ \ These
results are briefly summarized in the notation of this paper as
follows.

A convenient basis of $\text{\boldmath $\mathrm{H}$}^{\sigma }$
is the Gel'fand \cite{Gelfand} basis $|M\rangle $ that is derived
from the subgroup chain $\mathcal{U}( 1,n) \supset \mathcal{U}(
n) \supset \mathcal{U}( n-1) \supset ...\supset \mathcal{U}( 1)
$.\ \ This basis may be written explicitly as 
\[
\left. \left| M\right\rangle  =|\begin{array}{lllll}
 m_{1,n+1} & \ \ \ \ m_{2,n+1} & ...  & m_{n,n+1} & \ \ \ \ \ m_{n+1,n+1}
\\
   & m_{1,n} &  ... & \ \  & m_{n,n} \\
   &   &   &  ... &   \\
   &   & m_{1,1} &   &  
\end{array}\right\rangle  
\]

\noindent where the $m_{i,j}$ satisfy the inequalities
\[
m_{i,j}\geq m_{i+1,j},\ \ \ \ \ m_{i,j}\geq m_{i,j-1}\geq m_{i+1,j}\ \ .
\]

\noindent If you put the above inequalities into the triangular
form above, a simple pattern emerges. The integers $m_{i,n+1}$ label
the irreducible representations and the eigenvalues in this basis
of the Casimir operators of $\mathcal{U}( 1,n) $ are given in terms
of these quantities as is described shortly.\ \ The remainder of
the $m_{i,j}$ label states within the irreducible representation.

The full set of matrix elements of the representation
\begin{equation}
\left( \Sigma _{a,b}\right) _{\tilde{M},M}=\left\langle  \tilde{M}\right|
\sigma ^{\prime }( Z_{a,b}) \left| M\right\rangle  %
\label{sigma matrix elements}
\end{equation}

\noindent are determined by the matrix elements of\ \ $\sigma ^{\prime
}( Z_{a,a}) $, $\sigma ^{\prime }( Z_{a,a+1}) $ and $\sigma ^{\prime
}( Z_{a+1,a}) $. The remainder may be computed directly from these
using the Lie algebra relations.\ \ 

We need to convert between the indices $a,b=0,1...n$ used throughout
the paper and the $j,k..=1,...n+1$ that label the Gel'fand basis.
We therefore defined $\hat{k}=k$ for $k=1,...n$ and $\hat{k}=0$
for $k=n+1.$ Then, the matrix elements, as given in \cite{Maekawa},\cite{Ottoson}
are
\begin{equation}
\begin{array}{l}
 \sigma ^{\prime }( Z_{\hat{k},\hat{k}}) \left| M\right\rangle 
=\epsilon _{k,k}(  \operatorname*{\sum }\limits_{i=1}^{k}m_{k,i}-\operatorname*{\sum
}\limits_{i=1}^{k-1}m_{k-1,i}) \left| M\right\rangle  ,  \\
 \sigma ^{\prime }( Z_{\hat{k},\hat{k+1}}) \left| M\right\rangle
=\epsilon _{k,k+1}\operatorname*{\sum }\limits_{i=1}^{k}s_{k,i}(
M) \left| M+I_{k,i}\right\rangle  , \\
 \sigma ^{\prime }( Z_{\hat{k+1},\hat{k}}) \left| M\right\rangle
=\epsilon _{k+1,k}\operatorname*{\sum }\limits_{i=1}^{k}s_{k,i}(
M-I_{k,i}) \left| M-I_{k,i}\right\rangle  ,
\end{array}
\end{equation}

\noindent where $\epsilon _{k,j}=1$ for $j,k=1,..n$, $-1$ for $j=k=n+1$
and $i$ otherwise.\ \ $|I_{l,k}\rangle  $ is the state where all
the $m_{i,j}=0$ unless $i=l$ and $k=j$ in which case $m_{l,k}=1$.\ \ The\ \ \ $s_{k,j}$
are the functions
\begin{equation}
s_{k,j}( M) \doteq i\frac{\prod \limits_{i=1}^{k+1}r_{i,j}( k+1,0)
\ \ \prod \limits_{i=1}^{k-1}r_{i,j}( k-1,1) }{ \prod \limits_{i=1,i\neq
j}^{k-1}r_{i,j}( k,1) r_{i,j}( k,0)  }.%
\label{sigma rep f function}
\end{equation}

\noindent where
\[
r_{i,j}( k,a)  =\sqrt{m_{k,i}-m_{k,j}-i+j-a} 
\]

Now, for the compact case where the indices $a,b=0,1,..n$ are restricted
to the case $i,j=1,..n$, the representations are finite dimensional.
In the non-compact case, the indices $m_{i,k}$ labeling the states
are not generally finite dimensional. However, representations of\ \ $\mathcal{U}(
1,n) $ may be decomposed into infinite direct sums of the finite
dimensional $\mathcal{U}( n) $\ \ representations.\ \ In this case,
$m_{i,n}$ label the irreducible finite dimensional representations
of\ \ $\mathcal{U}( n)  $ and, again, an infinite number of these
finite dimensional irreducible representations of\ \ $\mathcal{U}(
n) $ appear in each irreducible representation of\ \ $\mathcal{U}(
1,n) $.\ \ 

A simple direct computation shows that the eigenvalue of the representation
of the first Casimir invariant\ \ $U$ is given by
\[
\sigma ^{\prime }( U) \left| \varphi \right\rangle  =\eta ^{a,b}\sigma
^{\prime }( Z_{a,b}) \left| \varphi \right\rangle  =d_{1}\left|
\varphi \right\rangle  
\]

\noindent where 
\begin{equation}
\begin{array}{rl}
 d_{1} & =\eta ^{a,b}( \Sigma _{a,b}) _{M,M}=\sum \limits_{a=0}^{n}\eta
^{a,a }m_{\hat{a},n+1} \\
  & =m_{1,n+1}+m_{2,n+1}.....m_{n,n+1}-m_{n+1,n+1}
\end{array}.%
\label{unitary casimir d in terms of m}
\end{equation}

\noindent Similarly, higher order Casimir eigenvalues $d_{\alpha
}$ may similarly be computed in terms of the $\{m_{k,n+1}\}$ with
$k=1,..n+1$.\ \ 
\section{Field equations of the quaplectic group}\label{c:37}\label{Section
Field Equations}
\subsection{The general Casimir field equations }\label{c:28}

The field equations are defined in $ ($2$ )$ to be 
\begin{equation}
\varrho ^{\prime }( C_{\alpha }) \left| \psi \right\rangle  =c_{\alpha
}\left| \psi \right\rangle   \mathrm{with} \left| \psi \right\rangle
\in \text{\boldmath $\mathrm{H}$}^{\varrho }\ \ , \alpha =1,2...N_{c}%
\label{field equation definition 2}
\end{equation}

\noindent where for $\mathcal{Q}( 1,n) $ $N_{c}=n+2$ and, in addition
to $C_{1}=I$,\ \ the $C_{\alpha }$ are defined in terms of the $W_{a,b}$
given in $ ($14$ )$ ,\ \ 
\begin{equation}
C_{2\beta }=\eta ^{a_{1},a_{2\beta }}...\eta ^{a_{2\beta -2},a_{2\beta
-1}}W_{a_{1},a_{2}}...W_{a_{2\beta -1},a_{2\beta }} ,%
\label{Casimir in terms of W}
\end{equation}

\noindent with $\beta =1,..n+1$. The representation $\varrho ^{\prime
}( W_{a,b}) $ of these generators is\ \ \ 
\begin{equation}
\varrho ^{\prime }( W_{a,b}) \left| \psi \right\rangle  =c \sigma
^{\prime }( Z_{a,b}) \left| \varphi \right\rangle  \otimes \left|
\phi \right\rangle  \oplus \left| \varphi \right\rangle  \otimes
a \xi ^{\prime }( A_{a}^{+}) \xi ^{\prime }( A_{b}^{-}) \left| \phi
\right\rangle  %
\label{representation of W}
\end{equation}

\noindent where $a= 1-\frac{c}{s}$.\ \ Again,\ \ $c=c_{1}$ is the
eigenvalue of the representation,\ \ \ 
\[
\varrho ^{\prime }( I) \left| \psi \right\rangle  =c\left| \psi
\right\rangle  .
\]

The representation of the general Casimir invariants $\varrho ^{\prime
}( C_{\alpha }) $ involve products of the $\varrho ^{\prime }( W_{a,b})
$ and therefore $\xi ^{\prime }( A_{a}^{\pm }) $. As the matrix
elements of the representations $\xi ^{\prime }( A_{a}^{\pm }) $
are given in $ ($36$ )$ in terms of the differential operators,
this appears to imply that the field equations will be higher order
differential equations.\ \ However, the Lie algebra may be used
to rearrange the terms defining the Casimir invariant operators
so that it may be established that the invariants will result in
no more than second order differential equations.\ \ Using the algebra
for the representation of the generators $ ($34$ )$, the expression
for the representation of the Casimir invariant operator $\varrho
^{\prime }( C_{4}) $ given by substituting $ ($48$ )$ and $ ($47$
)$ into $ ($46$ )$ may be rearranged into the form 
\begin{equation}
\begin{array}{l}
 \eta ^{a,c}\eta ^{b,d} \xi ^{\prime }( A_{a}^{-}) \sigma ^{\prime
}( Z_{c,d})  \xi ^{\prime }( A_{b}^{+}) \left| \psi \right\rangle
\\
 =\sum \limits_{\alpha =1}^{2}\sum \limits_{\kappa =0}^{2-\alpha
+1}f_{\kappa ,\alpha }^{2}( n+1,a,c_{\gamma })  \left( \sigma ^{\prime
}( D_{\alpha }) \right)  ^{\kappa }\left| \psi \right\rangle  .
\end{array}
\end{equation}

\noindent where the $D_{\alpha }$ are the Casimir invariants of
the unitary group given in $ ($17$ )$, $D_{1}=U$ and $D_{2}=\eta
^{a,b}\eta ^{c,d}Z_{a,d}Z_{b,c}$ and the co-efficient functions
depend on the dimension of the space, the constant $a= 1-\frac{c}{s}$
and the Casimir eigenvalues $ c_{\gamma }$, $\gamma =1,2,4$, given
in $ ($46$ )$ that are constants for each of the irreducible representations
of the quaplectic group. The explicit form of these co-efficient
functions for the forth order eigenvalue equation are
\begin{equation}
\begin{array}{l}
 f_{0,1}^{2}( n,a,c_{\gamma }) =\frac{1}{2}\left(  c n( 1+n) -\left(
\frac{n+1}{a}+n-1\right) c_{2}+\frac{c_{2}^{2}+c_{4}}{a c} \right)
, \\
 f_{1,1}^{2}( n,a,c_{\gamma }) =\frac{1}{2a}\left( \left( 1+a\right)
c( 1+n) -2c_{2}\right) , \\
 f_{2,1}^{2}( n,a,c_{\gamma }) =\frac{1}{2a}c, \\
 f_{2,2}^{2}( n,a,c_{\gamma }) =\frac{1}{2a}c.
\end{array}%
\label{Casimir 4 canonical functions}
\end{equation}

\noindent The corresponding functions for the representation of
the Casimir invariant $\varrho ^{\prime }( C_{2}) $\ \ $ ($15$ )$
are simply 
\begin{equation}
f_{0,1}^{1}( n,a,c_{\gamma }) =c_{2}/a ,\ \ \ f_{1,1}^{1}( n,a,c_{\gamma
}) =c/a%
\label{Casimir 1 canonical functions}
\end{equation}

Note that this equation is a second order equation in the operators
$\{A_{a}^{\pm }\}$.\ \ Similar expressions may be obtained for the
remaining\ \ eigenvalue equations $ ($46$ )$ for $\beta =1,2,..n+1$
and it may be verified by direct computation that, $ ($at least
up to $C_{8}$$ )$, these equations are also second order equations
in terms of the operators $\{A_{a}^{\pm }\}$.\ \ In fact, the equation
for $C_{2\beta }$ with $\beta =1,..n+1$ is given by
\begin{equation}
\begin{array}{l}
 \eta ^{a_{1},a_{2}}...\eta ^{a_{2\beta -1},a_{2\beta }}\xi ^{\prime
}( A_{a_{1}}^{-}) \sigma ^{\prime }( Z_{a_{2},a_{3}}) \times ..
\\
 \ \ \ \ \ .... \times \sigma ^{\prime }( Z_{a_{2\beta -2},a_{2\beta
-1}})  \xi ^{\prime }( A_{a_{2\beta }}^{+}) \left| \psi \right\rangle
\\
 =\sum \limits_{\alpha =1}^{\beta }\sum \limits_{\kappa =0}^{\beta
-\alpha +1}f_{\kappa ,\alpha }^{\beta }( n+1,a,c_{\gamma }) \left(
\sigma ^{\prime }( D_{\alpha }) \right)  ^{\kappa }\left| \psi \right\rangle
\end{array}%
\label{general casimir equation}
\end{equation}

\noindent with $\gamma =1,2,...2\beta $.

Furthermore, as from $ ($19$ )$ the $D_{\alpha }$ and the $C_{2\beta
}$ commute, the state $|\psi \rangle $ may be taken to be eigenfunctions
of the representations of $D_{\alpha }$,\ \ $\sigma ^{\prime }(
D_{\alpha }) |\psi \rangle =d_{\alpha }|\psi \rangle $ . The $d_{\alpha
}$ are not constants on the irreducible representation but rather
label states within the representation.\ \ The matrix elements of
the representations $\sigma ^{\prime }$ in the $|M\rangle $ basis
are defined in terms of the $\Sigma _{a,b}$\ \ by $ ($42-44$ )$
and we use this to define
\[
\Sigma ^{\beta }_{b,c}= \eta ^{a_{2},a_{3}}...\eta ^{a_{2\beta -2},a_{2\beta
-1}} \Sigma _{b,a_{2}}.. \Sigma _{a_{2\beta -1},c}.
\]

Finally, using\ \ $ ($36$ )$ to give the matrix elements of $\xi
^{\prime }( A_{a}^{\pm }) $ with respect to the $|x\rangle $ basis,
the general field equations may be written as the eigenvalue equation
\begin{equation}
\begin{array}{l}
 \Sigma ^{\beta }_{b,c},_{\tilde{M},M}\left( x^{b}-\frac{\partial
}{\partial x^{b}}\right) \left( x^{c}+\frac{\partial }{\partial
x^{c}}\right) \psi _{M}( x)  \\
 = \hat{f}^{\beta }( n,a,c_{\gamma },d_{\alpha }) \psi _{\tilde{M}}(
x) 
\end{array}%
\label{field equations d eigenvalue general expression}
\end{equation}

\noindent with $c=c_{1}$ and the functions $\hat{f}^{\beta }$\ \ given
by 
\[
\hat{f}^{\beta }( n,a,c_{\gamma },d_{\alpha }) =\operatorname*{\sum
}\limits_{\alpha =1}^{\beta }\operatorname*{\sum }\limits_{\kappa
=0}^{\beta -\alpha +1}f_{\kappa ,\alpha }^{\beta }( n+1,a,c_{\gamma
}) d_{\alpha } ^{\kappa }
\]

\noindent where $\alpha =1,...\beta $ and $\gamma =1,2,...2\beta
$.\ \ Thus, the field equations are a set of simultaneous second
order differential equations. The $\Sigma ^{\beta }_{b,c}$ are countably
infinite dimensional matrices defined in terms of the $m_{i,j}$,
$i,j\leq n$\ \ and the\ \ $d_{\alpha } $ are defined in terms of
the $m_{i,n+1}$ label the various $\sigma $ irreducible representations
of $\mathcal{U}( 1,n) $ that appear in each irreducible representation
of $\mathcal{Q}( 1,n) $.\ \ \ \ The $c_{\gamma }$ label the irreducible
representations of $\mathcal{Q}( 1,n) $.\ \ 
\subsection{The relativistic harmonic oscillator field equation}\label{c:41}

In the Poincar\'e case, different field equations result from different
representations of the little group. The Klein-Gordon equation resulted
from the trivial representation. In the quaplectic group case, the
trivial representation\ \ $\sigma ^{\prime }( Z_{a,b}) =0$ results
in field equations that are not invariant under the generators of
the Heisenberg group.\ \ 

Using the factorization $\mathcal{U}( 1,n) \simeq \mathcal{U}( 1)
\otimes \mathcal{S}\mathcal{U}( 1,n) $, consider the\ \ a representation
where the representation $\sigma ^{\prime }( U) $ of $\mathcal{U}(
1) $ generator is nontrivial but where the\ \ $\sigma ^{\prime }(
\hat{Z}_{a,b}) $ of $\mathcal{S}\mathcal{U}( 1,n) $ are trivial,
$\sigma ^{\prime }( \hat{Z}_{a,b}) =0$. The representations of $\mathcal{Q}(
1,n) $ therefore degenerate to the representations of the oscillator
group $\mathcal{O}s( n+1) \simeq \mathcal{U}( 1) \otimes _{s}\mathcal{H}(
n+1) $ that has only two independent Casimir invariant operators.\ \ $C_{1}=I$
is trivial and the quadratic field equation $ ($54$ )$, where the
$f_{\kappa ,\alpha }^{\beta }$ for $\beta =1$ is given in $ ($51$
)$,\ \ is just\ \ \ \ 
\[
  \eta ^{a,b}( x^{b}-\frac{\partial }{\partial x^{b}}) \left( x^{a}+\frac{\partial
}{\partial x^{a}}\right) =\frac{c_{2} +c\ \ d_{1}}{a}\psi ( x) .
\]

\noindent This may be\ \ rearranged to the familiar equation for
the relativistic oscillator 
\[
 \left( \eta ^{a,b}( x^{a}x^{b}-\frac{\partial ^{2}}{\partial x^{a}\partial
x^{b}}) -\frac{c d_{1}-c_{2}}{a}- \left( n-1\right) \right)  \psi
( x) =0.
\]

\noindent This has solutions such that $\operatorname*{\lim }\limits_{|x|\rightarrow
\infty } \psi ( x) =0$ only if\ \ \ \ 
\[
  \frac{1}{a }\left( c\ \ d_{1}-c_{2}\right) + \left( n-1\right)
=2k+n-1 
\]

\noindent for all $k\in \mathbb{Z}^{+}$ and gives the result that
$k\equiv d_{1}$ and $c=2 a$ with $c_{2}=0$.\ \ Using the definition
of $a= 1-\frac{c}{s}$,\ \ this results in 
\[
a=\frac{s }{2\left( s-1\right) } ,\ \ \ c_{1}=c=\frac{s }{\left(
s-1\right) }.
\]
\subsection{Field equations in coherent basis}\label{c:41}

The general field equations can be cast into a simpler form by considering
a coherent basis $|\eta _{K},M\rangle $ instead of a coordinate
basis $|x,M\rangle $. The matrix elements
\[
\Xi _{\tilde{K},K}^{a,b}=\left\langle  \eta _{\tilde{K}}\right|
\xi ^{\prime }( A_{a}^{+}) \xi ^{\prime }( A_{b}^{-}) \left| \eta
_{K}\right\rangle  =s\left\langle  \eta _{\tilde{K}}\right| \rho
^{\prime }( Z_{a,b}) \left| \eta _{K}\right\rangle  \ \ 
\]

\noindent are defined by $ ($4.38-39$ )$.\ \ The field general field
equation is then
\begin{equation}
\Sigma ^{\beta }_{a,b},_{\tilde{M},M}\Xi _{\tilde{K},K}^{a,b}\psi
_{M,K}^{k}= \hat{f}^{\beta }( n,a,c_{\gamma },d_{\alpha }) \psi
_{\tilde{M},\tilde{K}}^{k}.%
\label{field equations general expression}
\end{equation}

The Hilbert space $\text{\boldmath $\mathrm{H}$}^{\varrho }$ is
the direct product of countably infinite vector spaces $\text{\boldmath
$\mathrm{H}$}^{\varrho }=\text{\boldmath $\mathrm{H}$}^{\sigma }\otimes
\text{\boldmath $\mathrm{H}$}^{\xi }$.\ \ As noted in $ ($40-41$
)$ the Hilbert space decomposes into the direct sum of subspaces
$\text{\boldmath $\mathrm{H}$}_{k}^{\xi }$ invariant under\ \ $\Xi
^{a,b}$ and hence the label $k$ on the states that is invariant.
Define $\Xi ^{k, a,b}$ to be $\Xi ^{a,b}$ restricted to $\text{\boldmath
$\mathrm{H}$}_{k}^{\xi }$.\ \ \ Then,\ \ if we consider for a moment
the field equations of $\mathcal{Q}( n) $ by restricting the indices
$a,b=0,1,..n$ to $i,j=1,..n$, the field equations become
\begin{equation}
 \Sigma ^{\beta }_{i,j},_{\tilde{M},M}\Xi _{\tilde{K},K}^{k, i,j}\psi
_{M,K}^{k}= \hat{f}^{\beta }( n,a,c_{\gamma },d_{\alpha }) \psi
_{\tilde{M},\tilde{K}}^{k}.%
\label{compact field equations general expression}
\end{equation}

Now, for each $k$, the $\Xi ^{k, i,j}$ are finite dimensional matrices
of dimension $k$ with $\text{\boldmath $\mathrm{H}$}_{k}^{\xi }\simeq
\text{\boldmath $\mathrm{V}$}^{k}$.\ \ Likewise, as the representation
is now compact, the\ \ $\Sigma _{i,j}$ and hence $\Sigma ^{\beta
}_{i,j}$ are finite integer dimensional with dimension and so $\text{\boldmath
$\mathrm{H}$}^{\sigma }\simeq \text{\boldmath $\mathrm{V}$}^{\dim
\sigma }$.\ \ All of the quantities are defined and it reduces to
a finite matrix eigenvalue problem that is solvable as both the
$\Xi ^{k, i,j}$ and $\Sigma _{i,j}$ are Hermitian.\ \ These give
rise to a set of spinning harmonic oscillators as will be discussed
in a subsequent paper.

In the non-compact case of general interest, the matrices are countably
infinite. Never-the-less, these comments give reason to believe
that the solution of these field equations are tractable. 
\section{Discussion}\label{c:43}

We were led to the quaplectic group from the very basic goal of
obtaining a dynamical group that encompassed both the\ \ Poincar\'e
group and the Heisenberg group. One new physical assumption, the
Born orthogonal metric hypothesis, was adopted. There are no other
essentially new physical assumptions in this paper. 

The dynamical group framework has proven its effectiveness in the
Poincar\'e case.\ \ We correspondingly obtained the unitary representations
of the quaplectic group and the associated field equations that
are the eigenvalue equations for the Hermitian representations of
the Casimir invariants. 

The Poincar\'e\ \ group reduces in the limit of $c\rightarrow \infty
$ to the Galilei group of nonrelativistic mechanics. The Galilei
group acting on the position-time space leaves the time subspace
invariant. Thus, in the Galilei case we can speak of absolute time.
The Poincar\'e\ \ group does not have this invariant subspace and
consequently time is relative to the observer. The Poincar\'e\ \ group
{\itshape mixes} the position and time degrees of freedom.

The same phenomena occurs with the quaplectic group. The quaplectic
group acts on a nonabelian time-position-momentum-energy space.\ \ In
a companion paper in preparation \cite{Low2}, we show that in the
limit $b\rightarrow \infty $, the quaplectic group contracts effectively
to the Poincar\'e\ \ group.\ \ This contracted group, acting on
the full nonabelian space, leaves invariant\ \ the position-time
subspace. Thus, in this limit there is the concept of absolute position-time\ \ space,
or as it is more usually stated, {\itshape space-time}. However,
under the full quaplectic group, this is no longer the case. For
strongly interacting states, {\itshape space-time} is relative to
the observer and in general, of the degrees of freedom of the nonabelian
space {\itshape mix}. That is, space-time can be transformed into
energy and momentum and vice versa. 

The field equations of the quaplectic group may be explicitly determined.
The simplest equation in the set is the relativistic oscillator
that is the counterpart in this theory of the Klein-Gordon equation
of the Poincar\'e case.\ \ In the general case, the Hilbert space
$\text{\boldmath $\mathrm{H}$}^{\sigma }$ of the internal degrees
of freedom corresponding to the little group $\mathcal{U}( 1,n)
$ is generally infinite dimensional.\ \ This is unlike the compact
group where the little groups of physical interest are compact and
finite dimensional. 

The Hilbert space $\text{\boldmath $\mathrm{H}$}^{\xi }$ decomposes
into spaces\ \ $\text{\boldmath $\mathrm{H}$}_{k}^{\xi }$ that are
invariant under the $\rho $ representation of the generators of
the little group. This $\rho $ representation arises because the
normal subgroup $\mathcal{H}( n+1) $ is non abelian and has no counterpart
in the Poincar\'e case.\ \ 

How do we reconcile these infinite dimensional internal degrees
of freedom?\ \ The $\mathcal{U}( 1,n) $ representations may be decomposed
into an infinite ladder of finite dimensional $\mathcal{U}( n) $
representations. Preliminary results indicate that under the group
contractions, these ladders {\itshape break} so that each finite
dimensional representation that is a rung of the ladder of representations
defining the infinite dimensional irreducible\ \ $\mathcal{U}( 1,n)
$ representation becomes a $ ($finite$ )$ irreducible representation
of the contracted group.\ \ This requires further investigation
but we know that particle states seem to appear in ladders although
we have apparently probed only the first three rungs in the interactions
that are accessible. 

The theory bounds relative rates of change of momentum, force, in
addition to the usual rates of change of position, velocity. This
is embodied by the four distinct Poincar\'e subgroups with representations
of translation generators represented by the quad
\[
\begin{array}{lll}
 \varrho ^{\prime }( T)  & \leftrightarrow  & \varrho ^{\prime }(
Q_{i})  \\
 \updownarrow  &   & \updownarrow  \\
 \varrho ^{\prime }( P_{i})  & \leftrightarrow  & \varrho ^{\prime
}( E) 
\end{array}
\]

The Lorentz group of two of these Poincar\'e groups is the transformation
of the usual $ ($velocity$ )$ special relativity and the Lorentz
group of the remaining two Poincar\'e groups is the reciprocally
conjugate $ ($force$ )$ special relativity.\ \ \ The representation
of the translation generators of only one of the four Poincar\'e\ \ subgroups
can be simultaneously diagonalized, and therefore observed,\ \ in
this nonabelian space. These are the 4 faces of the quad.\ \ 

Should the quaplectic dynamical group have experimental basis, many
of the\ \ Poincar\'e\ \ group physical concepts become approximate
in the same manner that nonrelativistic, Galilei group Casimir invariants
are only an approximate limiting case.\ \ Mass and spin are no longer
the Casimir invariant eigenvalues labeling the irreducible representations.
Actions of the representations of the general quaplectic can transform
states with a given spin and mass into a state where these are different.
The effects of the quaplectic group may become physically significant
for strongly interacting systems where the relative forces between
particle states approach $b$. These effects become apparent at the
Planck scales $\lambda _{\alpha }$ defined in terms of the $c,b$
and $\hbar $ basic dimensional constants. $ ($All three of these
constants are required in the quaplectic group unlike the Poincar\'e\ \ group
for which only $c$ appears.$ )$ Depending on the value of $b$, these
effects may be difficult to access directly. 

A possible calculation that would provide a test of the quaplectic
symmetry may be obtained by noting that both free and interacting
particle states are states in the Hilbert space of the representations
of the quaplectic group.\ \ The Hilbert space of the representations
of the Poincar\'e group are only states for free stable particle
particles and not for interacting or decaying particles \cite{saller}.\ \ The
usual Poincar\'e\ \ group is one of the four Poincar\'e\ \ groups
that are subgroups of the quaplectic group. Therefore, we can consider
a reduction of $\mathcal{Q}( 1,n) $ with respect to $\mathcal{E}(
1,n) $ $ ($or its cover$ )$.\ \ \ Each irreducible representation
of the quaplectic group is expected to contain a number of unitary
irreducible representations of\ \ the Poincar\'e group labeled by
mass and spin $(\mu ,s)$. As the quaplectic group has discrete representations,
one would expect these Poincar\'e irreducible representations also
be discrete,\ \ $(\mu _{k},s_{k})$, $k=1,2,...$. . This would be
a mass-spin spectrum and determine a discrete set of mass values
$\mu _{k}$ for the free particle states in each unitary irreducible
representation of the quaplectic group.\ \ We have definitive experimental\ \ information
about such discrete spectrums in the low energy regime\ \ against
which to test this data.\ \ \ The generators that are not in the
Poincar\'e subalgebra enable transitions between irreducible Poincar\'e
particle states. That is, a free particle with one $ ($discrete$
)$ mass and spin is transformed into another.

The mathematical problem of determining this embedding and its labelling
is difficult. It has been solved for $\mathcal{S}\mathcal{O}( 3)
\subset \mathcal{U}( 3) $\cite{jarvis}.\ \ One approach may be to
extend these to $\mathcal{S}\mathcal{O}( 1,n+1) \subset \mathcal{U}(
1,n+1) $. It is well known that the $\mathcal{E}( 1,n) $ is a group
contraction limit of\ \ $\mathcal{S}\mathcal{O}( 1,n+1) $ and it
is likewise true that $\mathcal{Q}( 1,n) $ is a group contraction
limit of $\mathcal{U}( 1,n) $.\ \ Thus, it may be possible to obtain
the desired embedding using these group contractions if the general
embedding $\mathcal{S}\mathcal{O}( 1,n+1) \subset \mathcal{U}( 1,n+1)
$ can be solved.

The theory that has been discussed is a {\itshape special} or\ \ global
theory. That is, it is the counterpart of the special relativity
theory as apposed to the general relativistic theory. It should
be possible to create a more general theory by lifting the identification
of $\mathbb{Q}^{1,n}$ to the tangent space of a general, nonabelian
manifold with curvature and making the parameters local. Interestingly,
Schuller \cite{Schuller} proved a {\itshape no go} theorem for Hermitian
metrics but the extra $I U$ term required for it to be a Casimir
invariant of the quaplectic group causes this {\itshape no go} theorem
to not be applicable.\ \ 

The next steps in this research are to show how the quaplectic group
reduces to effectively the Poincar\'e case in the low interaction
limit.\ \ This mathematically is the group contraction in the limit
$b\rightarrow \infty $ just as the Poincar\'e group contracts to
the Galilei group that is a semi-direct product with an Euclidian
homogeneous group.\ \ This is being addressed in a follow-on paper
\cite{Low2}. A closely associated problem is to determine how the
field equations $ ($53$ )$, representations of the Casimir operators,\ \ reduce
in this limit to effectively the usual Klein-Gordon, Dirac, Maxwell
and so on of the Poincar\'e theory.\ \ Note the discrete {\itshape
spectrum} on the right hand side of these equations. A first step
in this difficult problem is to study the field equations of the
compact case $ ($55$ )$ to understand the spectrum of the {\itshape
spinning} oscillator. The full field equations with nontrivial representations
of the $\mathcal{S}\mathcal{U}( 1,n) $ involving relativistic spinning
oscillators \cite{Moshinsky},\cite{Quesne2},\cite{Lange} need to
be explicitly studied.\ \ Finally, the reduction or the quaplectic
group with respect to the Poincar\'e group would give insight into
the mass-spin spectrum as described above.

The author thanks Peter Jarvis and Young Kim for their support and
insightful discussions of this research.


\begin{thebibliography}{000}
\bibitem{born1} Born, M. $ ($1938$ )$. \textit{A suggestion for
unifying quantum theory and relativity}. \textit{Proc. Roy. Soc.
London}, \textbf{A165}, 291--302. \label{born1}
\bibitem{born2} Born, M. $ ($1949$ )$. \textit{Reciprocity Theory
of Elementary Particles}. \textit{Rev. Mod. Phys.}, \textbf{21},
463--473. \label{born2}
\bibitem{mackey} Mackey, G. W. $ ($1976$ )$. \textit{The theory
of unitary group representations}. Chicago: University of Chicago
Press.\label{mackey}
\bibitem{dawber} Eliott, J. P., \& Dawber, P. G. $ ($1984$ )$. \textit{Symmetry
in Physics: VII}. New York: Oxford University Press.\label{dawber}
\bibitem{Low} Low, S. G. $ ($2002$ )$. \textit{Representations of
the canonical group, $ ($the semi-direct product of the unitary
and Weyl-Heisenberg groups$ )$, acting as a dynamical group on noncommutative
extended phase space}. \textit{J. Phys. A}, \textbf{35}, 5711--5729.
\href{arXiv:math-ph/0101024}{arXiv:math-ph$ /$0101024} \label{Low}
\bibitem{Hall} Hall, B. C. $ ($2000$ )$. \textit{Lie Groups, Lie
Algebras, and Representations: An Elementary Introduction}. New
York: Springer.\label{Hall}
\bibitem{bohm} Bohm, A. $ ($1993$ )$. \textit{Relativistic and Nonrelativistic
Dynamical Groups. }. \textit{Found. Phys.}, \textbf{23}, 751--767.
\label{bohm}
\bibitem{folland} Folland, G. B. $ ($1989$ )$. \textit{Harmonic
Analysis on Phase Space}. Princeton: Princeton University Press.\label{folland}
\bibitem{quesne} Quesne, C. $ ($1988$ )$. \textit{Casimir Operators
of Semidirect Sum Lie Algebras}. \textit{J. Phys. A}, \textbf{21},
L321--L324. \label{quesne}
\bibitem{popov} Popov, V., \& Perelomov, A. $ ($1967$ )$. \textit{The
Casimir Operators of the Unitary Group}. \textit{Soviet J. Nucl.
Phys.}, \textbf{5}, 489--491. \label{popov}
\bibitem{cailiello} Cainiello, E. R. $ ($1981$ )$. \textit{Is there
a Maximum Acceleration}. \textit{Il. Nuovo Cim.}, \textbf{32}, 65--70.
\label{cailiello}
\bibitem{Feoli} Feoli, A. $ ($2003$ )$. \textit{Maximal acceleration
or maximal accelerations}. \textit{Int. J. Mod. Phys.}, \textbf{D12},
271--280.  \href{arXiv: gr-qc/0210038}{arXiv: gr-qc$ /$0210038}
\label{Feoli}
\bibitem{Schuller} Schuller, F. $ ($2002$ )$. \textit{Born Infeld
Kinematics}. \textit{Ann. Phys.}, \textbf{299}, 1--34.  \href{arXiv:hep-ph/0203079}{arXiv:hep-ph$
/$0203079} \label{Schuller}
\bibitem{Amelino} Amelino-Camelia, G. $ ($2004$ )$. \textit{Quantum
physics, the cosomological constant and Planck-scale phenomenology}.
\textit{Classical and Quantum Gravity}, \textbf{21}$ ($13$ )$, 3095--3110.
\href{arXiv: gr-qc/0312014}{arXiv: gr-qc$ /$0312014} \label{Amelino}
\bibitem{Major} Major, M. E. $ ($1977$ )$. \textit{The quantum mechanical
representations of the anisotropic harmonic oscillator group}. \textit{J.
Math. Phys.}, \textbf{18}, 1938--1943. \label{Major}
\bibitem{Wolf} Wolf, J. A. $ ($1975$ )$. \textit{Representations
of Certain Semidirect Product Groups}. \textit{J. Func. Anal.},
\textbf{19}, 339--372. \label{Wolf}
\bibitem{Kim} Kim, Y. S., \& Noz, M. E. $ ($1991$ )$. \textit{Phase
Space Picture of Quantum Mechanics: Group Theoretical Approach}.
Singapore: World Scientific.\label{Kim}
\bibitem{Maekawa} Maekawa, T. $ ($1991$ )$. \textit{On the unitary
irreducible representations of SO$ ($n,1$ )$ and SU$ ($n,1$ )$ in
the scalar product with the nonintertwining operator}. \textit{J.
Math. Phys.}, \textbf{32}, 1193--1202. \label{Maekawa}
\bibitem{Ottoson} Ottoson, U. $ ($1968$ )$. \textit{A Classification
of the Unitary Irreducible Representations of SU$ ($N,1$ )$,}. \textit{Commun.
Math. Phys.}, \textbf{10}, 114--131. \label{Ottoson}
\bibitem{Gitman} Gitman, D. M., Shelepin, A. L., \& Shelepin, A.
L. $ ($1993$ )$. \textit{Coherent states of SU$ ($l,1$ )$ groups}.
\textit{J. Phys. A: Math. Gen}, \textbf{26}, 7003--7018. \label{Gitman}
\bibitem{Gelfand} Gel'fand, I., \& Tsetlin, M. $ ($1950$ )$. \textit{Finite
dimensional representations of the group of unimodular matrices}.
\textit{Dokl. Akad. Nauk SSSR}, \textbf{71}$ ($1$ )$, 825--828.\label{Gelfand}
\bibitem{Low2} Low, S. G. $ ($2005$ )$. \textit{Quaplectic group
quantum mechanics: Poincar\'e\ \ limit}. in preparation\label{Low2}
\bibitem{saller} Saller, H. $ ($2005$ )$. \textit{The Hilbert Spaces
for Stable and Unstable\ \ Particles}. preprint \href{arXiv:hep-th/0501074v1}{arXiv:hep-th$
/$0501074v1} \label{saller}
\bibitem{jarvis} Jarvis, P. D. $ ($2004$ )$. \textit{Resolution
of the GL$ ($3$ )$ O$ ($3$ )$ state labelling problem via O$ ($3$
)$ -invariant Bethe subalgebra of the twisted Yangian}. preprint
\href{arXiv:hep-th/0411026v1}{arXiv:hep-th$ /$0411026v1} \label{jarvis}
\bibitem{Moshinsky} Moshinsky, M., \& Szczepaniak, A. $ ($1990$
)$. \textit{The Dirac Oscillator}. \textit{ J. Phys. A: Math. Gen.},
\textbf{22}, L817--L819. \label{Moshinsky}
\bibitem{Quesne2} Quesne, C. $ ($1990$ )$. \textit{Symmetry Lie
algebra of the Dirac oscillator}. \textit{J. Phys. A}, \textbf{23},
2263--2272. \label{Quesne2}
\bibitem{Lange} de Lange, O. L. $ ($1991$ )$. \textit{Algebraic
properties of the Dirac oscillator}. \textit{J. Phys. A}, \textbf{24},
667--677. \label{Lange}
\end{thebibliography}
\end{document}